\documentclass[12pt]{article}
\usepackage{pstricks}
\usepackage{color}
\usepackage{amssymb,amsmath,bm,bbm}
\usepackage{epsf}
\usepackage{epsfig}
\usepackage{afterpage}
\usepackage{longtable}
\usepackage{cite}
\usepackage{latexsym,mathrsfs,dsfont}
\usepackage{graphics}
\usepackage{graphicx,booktabs,array}
\usepackage{url}
\usepackage{paralist}
\usepackage{bbold}
\usepackage[colorlinks=true, urlcolor=blue]{hyperref}
\usepackage{appendix}
\usepackage{tikz-feynman}
\tikzfeynmanset{compat=1.0.0}
\newcommand{\nn}{\nonumber}

\newcommand{\be}{\begin{equation}}
\newcommand{\ee}{\end{equation}}
\newcommand{\bea}{\begin{eqnarray}}
\newcommand{\eea}{\end{eqnarray}}
\newcommand{\dd}{\displaystyle}
\numberwithin{equation}{section}
\begin{document}

\begin{flushright}
    {BARI-TH/23-749}
\end{flushright}

\medskip

\begin{center}
{\Large\bf
  \boldmath{ Dalitz  decays $D_{sJ}^{(*)} \to D_s^{(*)} \ell^+  \ell^- $}} 
\\[0.8 cm]
{\large P.~Colangelo$^a$, F.~De~Fazio$^a$, F.~Loparco$^a$ and N. Losacco$^{a,b}$
 \\[0.5 cm]}
{\small
$^a$
Istituto Nazionale di Fisica Nucleare, Sezione di Bari, Via Orabona 4,
I-70126 Bari, Italy \\[0.1 cm]
$^b$Dipartimento Interateneo di Fisica ``M. Merlin'', Universit\`a  e Politecnico di Bari, \\ via Orabona 4, 70126 Bari, Italy
}
\end{center}

\vskip0.5cm


\begin{abstract}
\noindent
The Dalitz decays  of the positive-parity $D_{sJ}^{(*)}$ charmed mesons, $D_{sJ}^{(*)} \to D_s^{(*)} \ell^+  \ell^-$ with $J=0,1,2$ and $\ell=e, \mu$,  are important processes  to investigate the nature of the $D_{sJ}^{(*)}$ states. We analyze the full set of  decays, considering the four lightest $D_{sJ}^{(*)}$ mesons as belonging to the heavy quark spin doublets  $\dd s_\ell^P=\frac{1}{2}^+$ and $\dd \frac{3}{2}^+$, with $s_\ell^P$ the spin-parity of the light degrees of freedom in mesons. The description implies relations among the observables in  various modes. We study the
decay distributions in the dilepton invariant mass squared and  the distributions in the angle between the charged lepton momentum and the momentum of the produced meson, which are expressed in terms of universal form factors and of effective strong couplings. Such measurements  
 are feasible   at present facilities.
\end{abstract}

\thispagestyle{empty}
\newpage

\section{Introduction}
The  Dalitz decays of the positive-parity mesons with open charm and strangeness, $D_{sJ}^{(*)}\to D_s^{(*)} \ell^+  \ell^-$ 
with $J=0,1,2$ and $\ell=e, \mu$, are of  interest since they can shed new light on the nature of the $D_{sJ}^{(*)}$  states. $D_{s0}^*(2317)$ and $D_{s1}^\prime(2460)$  show  puzzling features, already emerged  at their first observation by   BABAR and CLEO Collaborations \cite{BaBar:2003oey,CLEO:2003ggt}. In particular, their mass is below the $DK$ and $D^* K$ thresholds \cite{Colangelo:2003vg}.  Together with $D_{s1}(2536)$ and $D^*_{s2}(2573)$, they are the lightest  
 positive-parity mesons with charm and strangeness \cite{ParticleDataGroup:2022pth}. The classification of the four states in  two heavy quark spin doublets with 
 $\dd s_\ell^P=\frac{1}{2}^+$ and $\dd \frac{3}{2}^+$, with $s_\ell^P$ the spin-parity of the light degrees of freedom in the mesons, implies relations among various observables, namely among mass parameters and widths \cite{Colangelo:2004vu,Colangelo:2005gb}. Relations can also be established among  different Dalitz modes,  thanks to  the hadronic parameters common to the various decay amplitudes. Such relations can be experimentally verified, and this provides us with a new  way  to probe the nature of the states,  discriminating ordinary quark-antiquark mesons from  hadrons with large molecular/multiquark components.\footnote{~A molecular structure of $D_{s0}^*(2317)$  is advocated in Prelovsek's  talk given at Hadron 2023, Genoa.} In that respect, the $D_{sJ}^{(*)}$ Dalitz modes complement the information from the electric dipole radiative decays \cite{Colangelo:2005hv}. We have to say that, at present, the  description of the full spectrum of mesons with a single charm or beauty quark in terms of heavy quark spin doublets 
 naturally emerges from data \cite{Colangelo:2012xi}.

The   Dalitz amplitudes involve terms with the virtual photon coupled to the charm quark and to the light $\bar s$ quark.
For such terms we use the effective QCD theory based on the heavy quark expansion and on the hidden gauge symmetry, together with the vector meson dominance (VMD)\cite{Kroll:1967it} for the contribution related to the light (anti)quark.  The features of the effective theory are described in Sec.~\ref{sec:efftheory}.  In Sec.~\ref{sec:amplitudes} we collect the expressions of the decay amplitudes, which involve two form factors parametrizing the  matrix elements of the charm vector current with positive- and negative- parity mesons organized in spin doublets.  The effective strong couplings of the charmed mesons with  $\phi(1020)$ mesons  also appear in the amplitudes. The expressions of  double- and single-decay  distributions are  given in Sec.~\ref{sec:distributions}, with some functions collected in the Appendix.  In Sec.~\ref{sec:numerics} we discuss examples of decay distributions useful for a comparison with measurement. Then we conclude.

\section{Dynamics of mesons with a single heavy quark}\label{sec:efftheory}
In the heavy quark limit $m_Q \to \infty$, mesons comprising a single heavy quark can be classified in doublets of  $s_\ell^P$, the spin-parity of the light degrees of freedom (light quark and gluons). Indeed,  in that limit
the spin of the heavy quark decouples from the strong dynamics, and  the QCD Lagrangian displays heavy quark symmetry,  invariance under heavy quark spin transformations  and heavy quark flavour transformations.\footnote{~For reviews see \cite{Neubert:1993mb,Manohar:2000dt}.}  

The four positive-parity mesons corresponding to the $P-$wave states in the constituent quark model can be collected in two doublets with  $\dd s_\ell^P=\frac{1}{2}^+$ ($J^P=0^+,1^+$) and  $\dd s_\ell^P=\frac{3}{2}^+$ ($J^P=1^+,2^+$).
The pseudoscalar and vector mesons belong to the   $\dd s_\ell^P=\frac{1}{2}^-$ doublet. 
In the case of charm,  the mesons in the $\dd s_\ell^P=\frac{1}{2}^-$ doublet are $(D_a,D_a^*)$,   $a=u,d,s$ being a light flavour index,  the two positive-parity doublets comprise  $(D_{a0}^*,D_{a1}^\prime)$ and $(D_{a1},D_{a2}^*)$.
Each doublet is  described by  a $4\times4$ matrix:
\bea
H_a & =& \frac{1+{\rlap{v}/}}{2}[P_{a\mu}^*\gamma^\mu-P_a\gamma_5] \,   \hspace*{6.3cm}    (s_\ell^P=\frac{1}{2}^-)   \label{neg}   \\
S_a &=& \frac{1+{\rlap{v}/}}{2} \left[P_{1a}^{\prime \mu}\gamma_\mu\gamma_5-P_{0a}^*\right]   \hspace*{6cm}    (s_\ell^P=\frac{1}{2}^+)    \label{pos1} \\
T_a^\mu &=&\frac{1+{\rlap{v}/}}{2} \Bigg\{ P^{\mu\nu}_{2a}
\gamma_\nu - P_{1a\nu} \sqrt{3 \over 2} \gamma_5 \left[ g^{\mu \nu}-{1 \over 3} \gamma^\nu (\gamma^\mu-v^\mu) \right]\Bigg\}  \hspace*{0.8cm}     (s_\ell^P=\frac{3}{2}^+)   .
\label{pos2}
\eea
In \eqref{neg}-\eqref{pos2}  the  fields $P_{Ja}^{(*)}$ are normalized including a factor $\sqrt{m_{D_a^{(*)}}}$ and have dimension $3/2$.  $v$ is the heavy meson four-velocity, which is conserved in strong interaction processes.  

The low-energy Lagrangian describing the strong interactions of the heavy mesons and the light pseudoscalar and vector mesons can be constructed on the basis of the heavy quark symmetry, the chiral symmetry and the principle of hidden gauge invariance \cite{Wise:1992hn,Burdman:1992gh,Cho:1992gg,Yan:1992gz,Jenkins:1995vb,Casalbuoni:1996pg}.
The octet of light pseudoscalar mesons is introduced defining 
$\displaystyle \xi=e^{i {\cal M} \over f_\pi}$ and $\Sigma=\xi^2$.  The matrix ${\cal M}$
comprises the  $\pi$, $K$, and $\eta^{(8)}$ fields (the normalization corresponds to $f_{\pi}=132 \;$ MeV):
\begin{equation}
{\cal M}= \left(\begin{array}{ccc}
\sqrt{\frac{1}{2}}\pi^0+\sqrt{\frac{1}{6}}\eta^{(8)} & \pi^+ & K^+\\
\pi^- & -\sqrt{\frac{1}{2}}\pi^0+\sqrt{\frac{1}{6}}\eta^{(8)} & K^0\\
K^- & {\bar K}^0 &-\sqrt{\frac{2}{3}}\eta^{(8)}
\end{array}\right) \,\,\,\, . \label{pseudo-octet}
\end{equation}
The fields $\Sigma$ and $\xi$ transform  under the chiral group $SU(3)_L \times SU(3)_R$  as
 \bea
 \Sigma &\to& L\Sigma R^\dagger  \nn \\
  \xi (x) &\to& L\xi U^\dagger(x)=U(x) \xi R^\dagger  \,\,\, , \label{transf-xi} 
 \eea
 with $L (R)$ belonging to $SU(3)_{L(R)}$, and  $U(x)$  to the unbroken subgroup  $SU(3)_V$.
Vector and axial-vector currents can be constructed in terms of $\xi$,
 \begin{eqnarray}
{\cal V}_{\mu }&=&\frac{1}{2}\left(\xi^\dagger\partial_\mu \xi+\xi\partial_\mu \xi^\dagger\right) \;  \label{Vmu} \\
{\cal A}_{\mu }&=&\frac{i}{2}\left(\xi^\dagger\partial_\mu \xi-\xi \partial_\mu \xi^\dagger\right) \;  \label{Amu}
\end{eqnarray}
with  transformation properties
 \bea
{\cal A}_{\mu} &\to & U {\cal A}_{\mu} U^\dagger  \label{transf-A} \\
{\cal V}_{\mu} &\to & U {\cal V}_{\mu} U^\dagger +U \partial_\mu U^\dagger \,\,\, . \label{transf-V} 
\eea
The interactions of the heavy mesons with the light vector mesons can be constructed using the principle of hidden gauge symmetry\cite{Jenkins:1995vb,Casalbuoni:1996pg}. 
 The fields $\rho_\mu$ are introduced,
\be
 \rho_\mu=i \frac{g_V}{\sqrt{2}}{\hat \rho}_\mu 
 \label{rho}
 \ee
 in terms of the Hermitian fields of vector mesons
 \begin{equation}
{\hat \rho}_\mu= \left(\begin{array}{ccc}
\sqrt{\frac{1}{2}}\rho^0+\sqrt{\frac{1}{6}}\phi^{(8)} & \rho^+ & K^{*+}\\
\rho^- & -\sqrt{\frac{1}{2}}\rho^0+\sqrt{\frac{1}{6}}\phi^{(8)} & K^{*0}\\
K^{*-} & {\bar K}^{*0} &-\sqrt{\frac{2}{3}}\phi^{(8)}
\end{array}\right)_\mu \,\,\,\, . \label{matvecto1}
\end{equation}
The value   $g_V= 5.8$   is chosen to satisfy the Kawarabayashi-Suzuki-Riazuddin-Fayyazuddin relations \cite{Kawarabayashi:1966kd,Riazuddin:1966sw}.
 The  mesons $\omega$ and $\phi$ correspond to  the mixing between the flavour octet  $\phi^{(8)}$  in  (\ref{matvecto1}) and the flavour singlet component $\phi^{(0)}$, 
\bea
\phi&=&\sin \theta_V \phi^{(0)}-\cos \theta_V \phi^{(8)} \nn \\
\omega&=& \cos \theta_V \phi^{(0)}+\sin \theta_V \phi^{(8)} \,\, .
\eea
Flavour eigenstates $\phi_q=\displaystyle\frac{{\bar u}u+{\bar d}d}{\sqrt{2}}$ and $\phi_s={\bar s}s$ are obtained for  $\theta_V = {\rm arctan} \displaystyle\frac{1}{\sqrt{2}}$, and the observed  $\omega$ and $\phi$ are identified with  $\omega=\phi_q$ and $\phi=\phi_s$. 
Replacing $\displaystyle\frac{1}{\sqrt{3}}\phi^{(8)} =\sin \theta_V \phi^{(8)} \to \phi_q$, and $-\displaystyle\frac{2}{\sqrt{3}}\phi^{(8)} =-\cos \theta_V \phi^{(8)} \to \phi_s$,  the ideal $\omega-\phi$ mixing
(which is exact in the large $N_c$ limit  \cite{Jenkins:1995vb}) we have
 \begin{equation}
{\hat \rho}_\mu= \left(\begin{array}{ccc}
\sqrt{\frac{1}{2}}\rho^0+\sqrt{\frac{1}{2}}\omega & \rho^+ & K^{*+}\\
\rho^- & -\sqrt{\frac{1}{2}}\rho^0+\sqrt{\frac{1}{2}}\omega & K^{*0}\\
K^{*-} & {\bar K}^{*0} &\phi
\end{array}\right)_\mu \,\,\,\, . \label{matvector}
\end{equation}
The interactions of the positive- and negative-parity heavy quark   doublets with the light vector mesons,  of interest for our analysis, are described by the effective Lagrangian terms
\cite{Casalbuoni:1996pg,Campanella:2018xev}
\bea
{\cal L}_1^S&=&-g_1^S \, Tr \left[ {\bar H} { S} \gamma^\alpha ({\cal V}_\alpha-\rho_\alpha)\right] +{\rm H.c.}
\label{LHSV1} \\
{\cal L}_2^S&=& g_2^S \, \frac{1}{\Lambda}Tr \left[ {\bar H} { S}  \sigma^{\alpha \beta}{\cal F}_{\alpha \beta}\right]  + {\rm H.c.}  \label{LHSV2}
\eea
(with $\bar H = \gamma^0 H^\dagger \gamma^0$), and
\be 
{\cal L}_2^T=i \,h^T \, \frac{1}{\Lambda^2} Tr \left[ {\bar H} T_\mu \sigma^{ \alpha \beta} {\cal D}^\mu {\cal F}_{\alpha \beta}\right]  + {\rm H.c.} \,\, .  \label{LHTV2}
\ee
The  field strength ${\cal F}_{\alpha \beta}$ and the covariant derivative ${\cal D}_\mu$  are defined, respectively, as
 \bea
{\cal F}_{\mu \nu}&=&\partial_\mu \rho_\nu- \partial_\nu \rho_\mu+[\rho_\mu,\,\rho_\nu]  \nn \\
{\cal D}_\mu &=& \partial_\mu +  {\cal V}_\mu \,\, .
\eea
$g^S_{1,2}$ and $h^T$ are dimensionless low-energy  constants, and  $\Lambda$ is  a  low-energy scale set to $\Lambda=1$ GeV. 

We compute  the  $D_{sJ}^{(*)}$ Dalitz amplitudes using this formalism.

\section{ $D_{sJ}^{(*)}$ Dalitz amplitudes}\label{sec:amplitudes}
\begin{figure}[!t]
\begin{center}
\includegraphics[width = 0.80\textwidth]{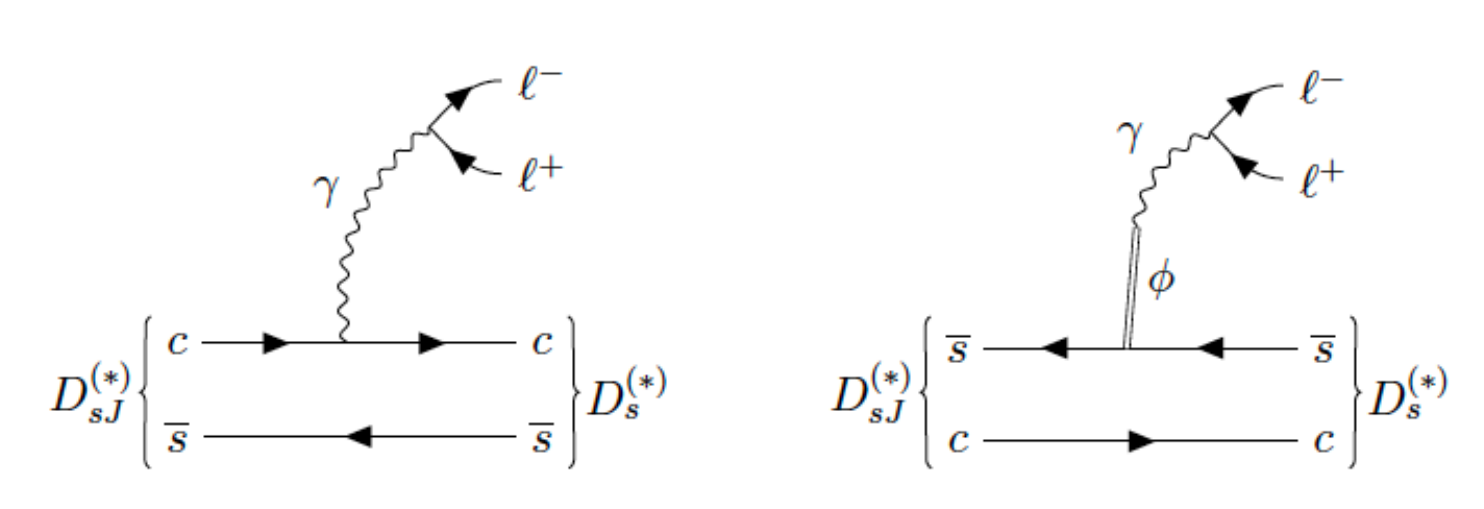}
\caption{\small  $D_{sJ}^{(*)} \to D_s^{(*)} \ell^+  \ell^-$ amplitude: photon coupled to the charm quark (left) and vector meson dominance  contribution for the photon coupled to the light  $\bar s$ (right).}\label{diag}
\end{center}
\end{figure}

The low-energy theory has been applied to the heavy vector meson magnetic dipole transitions   $D^*_a \to D_{a} \gamma $
 \cite{Cho:1992nt,Amundson:1992yp}. Here we focus on the electric dipole transitions of positive-parity heavy mesons, which contribute to the Dalitz amplitudes.
 The  $D_{sJ}^{(*)} \to D_s^{(*)} \ell^+  \ell^-$  amplitude reads
\be
{\cal A}(D_{sJ}^{(*)} (p^\prime)\to D_s^{(*)} (p) \ell^-(p_1)  \ell^+(p_2))=\langle D_s^{(*)}(p, \epsilon) |i J_\mu^{\rm em}|D_{sJ}^{(*)}(p^\prime, \epsilon^\prime) \rangle \frac{-i g^{\mu \nu}}{q^2}(-ie){\bar u}(p_1)\gamma_\nu v(p_2) ,
\label{amp}
\ee
where $q=p_1+p_2=p^\prime-p$  is the dilepton momentum. $\epsilon^\prime$ is the polarization vector of $D_{s1}^{(\prime)}$ or the polarization tensor of  $D^*_{s2}$,  and $\epsilon$ is the polarization vector of $D_{s}^*$.
The  relevant term of the  electromagnetic current is
\be
J_\mu^{\rm em}= e \left (\,e_c\,{\bar c}\gamma_\mu c + e_s\,{\bar s}\gamma_\mu s  \right) \,\, ,
\ee
with $e_c$ and $e_s$ the charm and strange quark electric charge, respectively (in units of $e$).
 The amplitude comprises two contributions  depicted  in Fig.~\ref{diag}, with the photon coupled to the charm quark and to the light antiquark $\bar s$.
 
 \vspace*{0.5cm}
 
For  the photon coupled to the charm quark,
the   ${\bar c} \Gamma c$ current (with $\Gamma$ a generic Dirac matrix)  matrix elements of the $S, H$ and $T,  H$ doublets can be computed in the effective theory using the trace formalism \cite{Falk:1991nq}. They involve the $\tau_{1/2}(w)$ (for $S \to H$) and $\tau_{3/2}(w)$ (for $T \to H$)  universal functions  \cite{Isgur:1990jf}:
\bea
\langle H(v)| {\bar c} \Gamma c |S(v^\prime) \rangle &=&- \tau_{1/2}(w)\, {\rm Tr}[{\bar H}(v) \,\Gamma \,S(v^\prime) ] \nn\\
\langle H(v)| {\bar c} \Gamma c |T(v^\prime) \rangle &=& -\tau_{3/2}(w)\, {\rm Tr}[{\bar H}(v)\, \Gamma  \, v_\mu \,T^\mu(v^\prime) ]  \,\, ,
\eea
with $w$   the product of meson four-velocities $w=v^\prime \cdot v =\displaystyle\frac{m_{D_{sJ}^{(*)}}^2+m_{D_s^{(*)}}^2-q^2}{2 m_{D_{sJ}^{(*)}}m_{D_s^{(*)}}}$.
The  $D_{sJ}^{(*)} \to D_s^{(*)}$ matrix elements of the charm vector current read: 
\bea
\langle D_s^*(m_{D_s^*} v,\, \epsilon)|\bar c  \gamma_\mu c |D_{s0}^* (m_{D_{s0}^*} v^\prime)\rangle &=&\tau_{1/2}(w)\sqrt{m_{D_{s0}^*} m_{D_s^* }}\left[-(w-1)\epsilon^*_\mu+
(\epsilon^* \cdot v^\prime) v_\mu \right]\,\, \,\,\,\\
\langle D_s(m_{D_s} v)|\bar c  \gamma_\mu c |D_{s1}^\prime (m_{D_{s1}^\prime} v^\prime, \, \epsilon^\prime)\rangle &=&\tau_{1/2}(w)\sqrt{m_{D_{s1}^\prime} m_{D_s} }\left[(w-1)\epsilon^\prime_\mu-
(\epsilon^\prime \cdot v) v^\prime_\mu \right]\,\, \,\,\,\\
\langle D_s^*(m_{D_s^*} v,\, \epsilon)| \bar c  \gamma_\mu c |D_{s1}^\prime (m_{D_{s1}^\prime} v^\prime, \, \epsilon^\prime)\rangle &=&\tau_{1/2}(w)\sqrt{m_{D_{s1}^\prime} m_{D_s^*} }\,i\,\left(\epsilon_{\alpha \beta \tau \mu} \epsilon^{\prime \alpha} \epsilon^{*\beta} v^\tau -\epsilon_{\alpha \beta \tau \mu} \epsilon^{\prime \alpha} \epsilon^{*\beta} v^{\prime \tau} \right)  \nn \\ \,\,\,\,\, \\
\langle D_s(m_{D_s} v)| \bar c  \gamma_\mu c |D_{s1} (m_{D_{s1}} v^\prime, \, \epsilon^\prime)\rangle &=&\tau_{3/2}(w)\sqrt{m_{D_{s1}}m_{D_s} }\nn \\
&\times&\frac{1}{\sqrt{6}}\left\{ (w^2-1)\epsilon^\prime_\mu+(\epsilon^\prime \cdot v)  \left(3v_\mu-(w-2)v^\prime_\mu \right) \right\}  \\
\langle D_s^*(m_{D_s^*} v,\, \epsilon)|\bar c  \gamma_\mu c |D_{s1} (m_{D_{s1}} v^\prime, \, \epsilon^\prime)\rangle &=&\tau_{3/2}(w)\left(-\frac{i}{\sqrt{6}}\right)\sqrt{m_{D_{s1}}m_{D_s^*} } \nn \\&& \hskip -6cm
\times \left\{(w-1)\left(\epsilon_{\alpha \beta \tau \mu} \epsilon^{\prime \alpha} \epsilon^{*\beta} v^\tau +\epsilon_{\alpha \beta \tau \mu} \epsilon^{\prime \alpha} \epsilon^{*\beta} v^{\prime \tau}\right)+(\epsilon^\prime \cdot v) \epsilon_{\alpha \beta \tau \mu} \epsilon^{* \alpha} v^\beta v^{\prime \tau}
-2 \epsilon_{\alpha \beta \tau \sigma} \epsilon^{\prime \alpha} \epsilon^{*\beta}v^\tau v^{\prime \sigma}v_\mu \right\} \nn \\ \\
\langle D_s(m_{D_s} v)| \bar c  \gamma_\mu c |D_{s2}^* (m_{D_{s2}^*} v^\prime, \, \epsilon^\prime)\rangle &=&\tau_{3/2}(w)\, i\,\sqrt{m_{D_{s2}^*}m_{D_s} }
\,\epsilon_{\mu \nu \alpha \beta} \epsilon^{\prime  \tau \nu}v_\tau v^\alpha v^{\prime \beta}  \\
\langle D_s^*(m_{D_s^*} v,\, \epsilon)| \bar c  \gamma_\mu c |D_{s2}^* (m_{D_{s2}^*} v^\prime, \, \epsilon^\prime)\rangle &=&\tau_{3/2}(w)\,\sqrt{m_{D_{s2}^*}m_{D_s^*} }\nn \\
&\times&\Big\{(\epsilon^* \cdot v^\prime) \epsilon^\prime_{\tau \mu}v^\tau+\epsilon^\prime_{\tau \sigma} v^\tau v^\sigma \epsilon_\mu^*-\epsilon^\prime_{\tau \sigma} v^\tau \epsilon^{*\sigma}(v_\mu+v^\prime_\mu) \Big\} .
\eea
$\langle D_s |\bar c  \gamma_\mu c |D_{s0}^* \rangle$ vanishes.

The term with the photon coupled to $\bar s$  involves a long-distance contribution that can be computed using  the vector meson dominance  \cite{Colangelo:1993zq,Colangelo:1994jc}.
The matrix elements of the strange quark vector current  are written, neglecting the $\phi$ width, as
\be
\langle D_s^{(*)} (p)|{\bar s}\gamma_\mu s|D_{sJ}^{(*)}(p^\prime)\rangle=\langle D_s^{(*)}(p)\phi(q, \eta)|D_{sJ}^{(*)}(p^\prime) \rangle \,\frac{i}{q^2-m_\phi^2}\langle 0 |{\bar s}\gamma_\mu s| \phi(q,\eta) \rangle \label{eq:amps}
\ee
 with
\be
\langle 0 |{\bar s}\gamma_\mu s|\phi(q,\eta) \rangle=m_\phi f_\phi \eta_\mu \,\,.
\ee
The decay constant $f_\phi$,  set by the measurement of  $\Gamma(\phi \to \mu^+ \mu^-)$,  is quoted in Table \ref{tab:input}.
The strong matrix elements of positive and negative parity charmed states with $\phi(1020)$ 
\be
\langle D_s^{(*)}(p)\phi(q, \eta)|D_{sJ}^{(*)}(p^\prime) \rangle={\cal A}_\alpha^{ D_{sJ}^{(*)} D_s^{(*)} \phi} \eta^{*\alpha}
\ee
 can be obtained from the low-energy Lagrangians \eqref{LHSV1},  \eqref{LHSV2} and  \eqref{LHTV2} 
 \cite{Casalbuoni:1996pg,Campanella:2018xev}.
Hence, the  elements in \eqref{eq:amps} are written as
\be
\langle D_s^{(*)} (p)|{\bar s}\gamma_\mu s|D_{sJ}^{(*)}(p^\prime)\rangle={\cal A}_\alpha^{ D_{sJ}^{(*)} D_s^{(*)} \phi}   \,\frac{i}{q^2-m_\phi^2}m_\phi f_\phi \left( -g_\mu^\alpha+\frac{q^\alpha q_\mu}{q^2} \right) \,\,, 
\ee
with the following  amplitudes ${\cal A}_\alpha^{ D_{sJ}^{(*)}D_s^{(*)}\phi}$:
\bea
{\cal A}_\alpha^{ D_{s0}^*D_s^*\phi}&=&\frac{i g_V}{\sqrt{2}}\sqrt{\frac{m_{D_{s0}^*}}{m_{D^*_s}}} \Big\{
m_{D_{s0}^*}\left(g_1^S+2(m_{D_{s0}^*}+m_{D^*_s})\frac{g_2^S}{\Lambda}\right)(\epsilon^* \cdot v^\prime) \, v^\prime_\alpha\nn \\
 &-&
m_{D^*_s}(w+1)\left(g_1^S +2(m_{D_{s0}^*}-m_{D^*_s})\frac{g_2^S}{\Lambda}\right)\epsilon^*_\alpha\Big\} \,\, , 
\eea
\bea
{\cal A}_\alpha^{ D_{s1}^\prime D_s \phi}&=&\frac{i g_V}{\sqrt{2}}\sqrt{m_{D_{s1}^\prime}m_{D_s}}\Big\{
\left(g_1^S -2(m_{D_{s1}^\prime}+m_{D_s})\frac{g_2^S}{\Lambda}\right)(\epsilon^\prime \cdot v) \, v^\prime_\alpha\nn \\
 &-&(w+1)\left(g_1^S+2(m_{D_{s1}^\prime}-m_{D_s})\frac{g_2^S}{\Lambda}\right)\epsilon^\prime_\alpha\Big\} \,\, , \\
{\cal A}_\alpha^{ D_{s1}^\prime D_s^* \phi}&=&\frac{ g_V}{\sqrt{2}}\sqrt{m_{D_{s1}^\prime}m_{D_s^*}}\Big\{
\big[g_1^S+2\frac{g_2^S}{\Lambda}\left(m_{D_{s1}^\prime}-m_{D_s^*}(1+2w) \right)\big]\epsilon_{\alpha \beta \sigma \tau}\epsilon^{*\beta}\epsilon^{\prime \sigma}v^\tau \nn \\
&+&\big[g_1^S+2\frac{g_2^S}{\Lambda}\left(m_{D_{s1}^\prime}+m_{D_s^*} \right)\big]\epsilon_{\alpha \beta \sigma \tau}\epsilon^{*\beta}\epsilon^{\prime \sigma}v^{\prime \tau} 
-4\frac{g_2^S}{\Lambda}m_{D_s^*} (\epsilon^\prime \cdot v)\epsilon_{\alpha \beta \sigma \tau}\epsilon^{*\beta}v^\sigma v^{\prime \tau}
\Big\}  , \quad 
\eea
\bea
{\cal A}_\alpha^{ D_{s1}D_s \phi}&=&\frac{i g_V h_T}{\sqrt{3}\Lambda^2}\sqrt{m_{D_{s1}}m_{D_s}}\Big\{-m_{D_s}(m_{D_{s1}}+m_{D_s})(w^2-1)\epsilon^\prime_\alpha  \nn \\
&+&\Big[m_{D_{s1}}^2+m_{D_s}^2(w+2)-m_{D_s}m_{D_{s1}}(1+3w)\Big](\epsilon^\prime \cdot v) v^\prime_\alpha
\Big\} \,\, ,\\
{\cal A}_\alpha^{ D_{s1}D_s^* \phi}&=&\frac{g_V h_T}{\sqrt{3}\Lambda^2}\sqrt{m_{D_{s1}}m_{D_s^*}}\,m_{D_s^*}\Big\{(m_{D_{s1}}-m_{D_s^*})(w+1)
\epsilon_{\alpha \beta \sigma \tau}\epsilon^{*\beta}\epsilon^{\prime \sigma}v^{\prime \tau} \nn \\
&+&(m_{D_{s1}}+(2w+3)m_{D_s^*})(\epsilon^\prime \cdot v)
\epsilon_{\alpha \beta \sigma \tau}\epsilon^{*\beta}v^\sigma v^{\prime \tau}  \nn \\
&-&(m_{D_{s1}}-(2w-1)m_{D_s^*})(w+1)\epsilon_{\alpha \beta \sigma \tau}\epsilon^{*\beta}\epsilon^{\prime \sigma}v^\tau \Big\}  \,\, ,
\eea
\bea
{\cal A}_\alpha^{ D_{s2}^*D_s \phi}&=&\frac{\sqrt{2} g_V h_T}{\Lambda^2}\sqrt{m_{D_{s2}^*}m_{D_s}}m_{D_s}(m_{D_{s2}^*}+m_{D_s})\epsilon_{\alpha \beta \sigma \tau}v^\beta v^{\prime \sigma} \epsilon^{\prime \nu \tau} v_\nu \,\, ,\\
{\cal A}_\alpha^{ D_{s2}^*D_s^* \phi}&=&\frac{i \sqrt{2} g_V h_T}{\Lambda^2}\sqrt{m_{D_{s2}^*}m_{D_s^*}}\Big\{(m_{D_{s2}^*}+m_{D_s^*})m_{D_s^*}(\epsilon^* \cdot v^\prime)\epsilon^{\prime \tau \alpha}v_\tau \nn \\
&-&(m_{D_{s2}^*}^2+m_{D_s^*}^2-2w\,m_{D_{s2}^*}m_{D_s^*})\epsilon^{\prime \tau \sigma}v_\tau  \epsilon^{*\sigma} v^\prime_\alpha  \nn \\
&+&m_{D_s^*}\epsilon^{\prime \tau \sigma}v_\tau  v_\sigma [(m_{D_{s2}^*}+m_{D_s^*}) \epsilon^*_\alpha-2m_{D_{s2}^*}(\epsilon^* \cdot v^\prime)v^\prime_\alpha]\Big\} .\qquad 
\eea

\section{Decay distributions}\label{sec:distributions}

The previous amplitudes allow us to compute the various decay distributions. 
We define the functions appearing in the distributions:
\bea
g(q^2) &=&\frac{ \alpha^2 }{12 \pi q^6}|\vec p| \sqrt{1-\frac{4m_\ell^2}{q^2}} \nn \\
 f(q^2)&=&\frac{f_\phi g_V m_\phi}{q^2-m_\phi^2 } \,\,\, ,
\eea
with 
$\dd |\vec p|=\frac{\lambda^{1/2}(m_{D_{sJ}^{(*)}}^2,m_{D_s^{(*)}}^2,q^2)}{2m_{D_{sJ}^{(*)}}}$
the modulus of  the $D_s^{(*)}$ three-momentum  in the   decaying particle rest frame.

\subsection{ $D_{sJ}^{(*)} \to D_s \ell^+  \ell^-$  distributions}

For $D_{sJ}^{(*)} \to D_s \ell^+  \ell^-$ the double differential distribution  in $q^2$ and $\cos \theta$,  $\theta$ being the angle between the charged lepton momentum and the $D_s$ momentum,  is written as
\begin{equation}
\frac{d^2 \Gamma}{d q^2 \, d \cos \theta} = \left( \frac{d^2 \Gamma}{d q^2 \, d \cos \theta} \right)_c + \left( \frac{d^2 \Gamma}{d q^2 \, d \cos \theta} \right)_s + \left( \frac{d^2 \Gamma}{d q^2 \, d \cos \theta} \right)_\text{int} \;.  \label{eq:distr}
\end{equation}
The subscripts $c$  and $s$ refer to the photon coupled to the charm and strange quark, the last term is  the interference.
Each term in \eqref{eq:distr} has the form
\begin{equation}
\left( \frac{d^2 \Gamma}{d q^2 \, d \cos \theta} \right)_X = A_X(q^2) + B_X(q^2) \, \cos^2 \theta \,\,  \label{eq:doubled}
\end{equation}
with $X = \{ c, s, \text{int} \}$. The $q^2$ distribution  is obtained integrating over $\cos \theta$,
\be
\frac{d \Gamma}{dq^2}=
\left(\frac{d \Gamma}{dq^2}\right)_c+\left(\frac{d \Gamma}{dq^2}\right)_s+\left(\frac{d \Gamma}{dq^2}\right)_{\rm int} . \label{eq:singled}
\ee
For the various processes, the expressions of the functions $A_X(q^2)$ and $B_X(q^2)$  in \eqref{eq:doubled}  are in the Appendix. 
The distributions in the lepton pair invariant mass squared are listed below.

\vspace{1cm}
\noindent {\fbox{$ \bf  {D_{s1}^\prime \to D_s \ell^+  \ell^-}$}}
\bea
\left(\frac{d \Gamma}{dq^2}\right)_c&=&g(q^2) [\tau_{1/2}(q^2)]^2\frac{2[(m_{D_{s1}^{\prime}}-m_{D_s})^2-q^2]^2}{27 \, m_{D_{s1}^\prime}^3\,m_{D_s}}
[(m_{D_{s1}^{\prime}}+m_{D_s})^2+2q^2](2m_\ell^2+q^2)  \qquad \\
\left(\frac{d \Gamma}{dq^2}\right)_s&=&\frac{g(q^2)\, f^2(q^2)}{108 \, m_{D_{s1}^{\prime}}^3m_{D_s}}
\big[(m_{D_{s1}^{\prime}}+m_{D_s})^2-q^2 \big]^2 (2m_\ell^2+q^2) \nn \\
&\times&
\Bigg\{(g_1^S)^2\Big[(m_{D_{s1}^{\prime}}-m_{D_s})^2\,+2q^2 \Big]
+\frac{12g_1^S\,g_2^S\,q^2}{\Lambda}(m_{D_{s1}^{\prime}}-m_{D_s}) \nn \\
&+&\frac{4(g_2^S)^2\,q^2}{\Lambda^2}\Big[2(m_{D_{s1}^{\prime}}-m_{D_s})^2+ q^2 \Big] \Bigg\} \\
\left(\frac{d \Gamma}{dq^2}\right)_{\rm int}&=&\frac{g(q^2) f(q^2)[\tau_{1/2}(q^2)]}{27 \, m_{D_{s1}^{\prime}}^3m_{D_s}}\sqrt{2}\lambda(m_{D_{s1}^{\prime}}^2,m_{D_s}^2,q^2)(2m_\ell^2+q^2) \nn \\
&\times &
\Bigg\{g_1^S\big[m_{D_{s1}^{\prime}}^2-m_{D_s}^2+2q^2\big]
+2\frac{g_2^S \, q^2}{\Lambda}\big[3m_{D_{s1}^{\prime}}-m_{D_s}\big]\Bigg\} \,\,\, .
\eea

\noindent {\fbox{$\bf D_{s1} \to D_s \ell^+  \ell^-$}}
\bea
\left(\frac{d \Gamma}{dq^2}\right)_c&=&g(q^2) [\tau_{3/2}(q^2)]^2\frac{\lambda^2(m_{D_{s1}}^2,m_{D_s}^2,q^2)}{162 \, m_{D_{s1}}^5\,m_{D_s}^3}
[2(m_{D_{s1}}+m_{D_s})^2+q^2](2m_\ell^2+q^2) \\
%
\left(\frac{d \Gamma}{dq^2}\right)_s&=&g(q^2)\,f^2(q^2)\frac{h_T^2 q^2}{\Lambda^4 \,324 \, m_{D_{s1}}^5m_{D_s}}
\lambda^2(m_{D_{s1}}^2,m_{D_s}^2,q^2) [(m_{D_{s1}}+m_{D_s})^2+2q^2](2m_\ell^2+q^2)\nn\\ \\
%
\left(\frac{d \Gamma}{dq^2}\right)_{\rm int}&=&-g(q^2)\, f(q^2) \tau_{3/2}(q^2)
\frac{h_T q^2}{\Lambda^2 \,81\sqrt{2}m_{D_{s1}}^5m_{D_s}^2}
\lambda^2(m_{D_{s1}}^2,m_{D_s}^2,q^2) (2m_\ell^2+q^2) (m_{D_{s1}}+m_{D_s}) \,\, . \nn \\
\eea

\noindent {\fbox{ $ \bf D_{s2}^* \to D_s \ell^+  \ell^-$}}
\bea
\left(\frac{d \Gamma}{dq^2}\right)_c&=&g(q^2) [\tau_{3/2}(q^2)]^2\frac{\lambda^2(m_{D_{s2}^*}^2,m_{D_s}^2,q^2)}{90\, m_{D_{s2}^*}^5\,m_{D_s}^3} q^2 (2m_\ell^2+q^2) \\
\left(\frac{d \Gamma}{dq^2}\right)_s&=&g(q^2)\, f^2(q^2) \frac{h_T^2 q^2}{\Lambda^4 \,180 \, m_{D_{s2}^*}^5m_{D_s}}
\lambda^2(m_{D_{s2}^*}^2,m_{D_s}^2,q^2) (m_{D_{s2}^*}+m_{D_s})^2(2m_\ell^2+q^2) \\
\left(\frac{d \Gamma}{dq^2}\right)_{\rm int}&=&g(q^2)\, f(q^2) \tau_{3/2}(q^2)
\frac{h_T q^2}{\Lambda^2 \,45\sqrt{2}m_{D_{s2}^*}^5m_{D_s}^2}
\lambda^2(m_{D_{s2}^*}^2,m_{D_s}^2,q^2) \, (2m_\ell^2+q^2) (m_{D_{s2}^*}+m_{D_s})  . \nn \\
\eea

\subsection{$D_{sJ}^{(*)} \to D_s^* \ell^+  \ell^-$ distributions }
For $D_{sJ}^{(*)} \to D_s^* \ell^+  \ell^-$ we consider the double distributions in $q^2$ and in the angle $\theta$ between the momentum of the charged lepton and the momentum of $D_s^*$. The functions $A_X(q^2)$ and $B_X(q^2)$ are in the Appendix, and the distributions in the invariant mass squared of the lepton pair are given below. 

\vspace*{1cm}
\noindent {\fbox{ $\bf D_{s0}^*\to D_s^* \, \ell^+ \,  \ell^-$}}

This process is kinematically allowed only for $\ell=e$. The  terms in the $q^2$ distribution  \eqref{eq:singled} read:
\bea
\left(\frac{d \Gamma}{dq^2}\right)_c  &=&  g(q^2) \, \tau_{1/2}(q^2) \, \frac{2 \, [ ( m_{D_{s0}^*} - m_{D_s^*} )^2 - q^2 ]^2}{9 \, m_{D_{s0}^*}^3\,m_{D_s^*}} \, [ (m_{D_{s0}^*} + m_{D_s^*} )^2 + 2 \, q^2 ] \, ( 2 \, m_\ell^2 + q^2 ) \\
%
\left(\frac{d \Gamma}{dq^2}\right)_s & = & \frac{g(q^2)\,f^2(q^2) }{36 \, m_{D_{s0}^*}^3 m_{D_s^*}}
\big[(m_{D_{s0}^*}+m_{D_s^*})^2-q^2 \big]^2 (2m_\ell^2+q^2)
 \Bigg\{(g_1^S)^2\Big[(m_{D_{s0}^*}-m_{D_s^*})^2\,+2q^2 \Big] \nn \\
&+&\frac{12g_1^S\,g_2^S\,q^2}{\Lambda}(m_{D_{s0}^*}-m_{D_s^*})
+\frac{4(g_2^S)^2\,q^2}{\Lambda^2}\Big[2(m_{D_{s0}^*}-m_{D_s^*})^2+ q^2 \Big] \Bigg\}\\
%
\left(\frac{d \Gamma}{dq^2}\right)_{\rm int}&=&\frac{g(q^2) f(q^2)[\tau_{1/2}(q^2)]}{9 \, m_{D_{s0}^*}^3m_{D_s^*}}\sqrt{2}\lambda(m_{D_{s0}^*}^2,m_{D_s^*}^2,q^2)(2m_\ell^2+q^2) \nn \\
&\times & \Bigg\{g_1^S\big[m_{D_{s0}^*}^2-m_{D_s^*}^2-2q^2\big]
-2\frac{g_2^S \, q^2}{\Lambda}\big[m_{D_{s0}^*}-3m_{D_s^*}\big]\Bigg\} \,\, .
\eea

\noindent {\fbox {$\bf D_{s1}^\prime\to D_s^* \ell^+  \ell^-$}}
\bea
\left(\frac{d \Gamma}{dq^2}\right)_c&=&g(q^2) [\tau_{1/2}(q^2)]^2\frac{4[(m_{D_{s1}^{\prime}}-m_{D_s^*})^2-q^2]^2}{27 \, m_{D_{s1}^\prime}^3\,m_{D_s^*}}[(m_{D_{s1}^{\prime}}+m_{D_s^*})^2+2q^2](2m_\ell^2+q^2) \nn \\
%
\left(\frac{d \Gamma}{dq^2}\right)_s&=&\frac{g(q^2)\, f^2(q^2)}{54 \, m_{D_{s1}^{\prime}}^3m_{D_s^*}}
\big[(m_{D_{s1}^{\prime}}+m_{D_s^*})^2-q^2 \big]^2(2m_\ell^2+q^2)  \nn \\
&\times&
\Bigg\{(g_1^S)^2\Big[(m_{D_{s1}^{\prime}}-m_{D_s^*})^2\,+2q^2 \Big]
+\frac{12 g_1^S\,g_2^S\,q^2}{\Lambda}(m_{D_{s1}^{\prime}}-m_{D_s^*}) \nn \\
&+&\frac{4(g_2^S)^2\,q^2}{\Lambda^2}\Big[2(m_{D_{s1}^{\prime}}-m_{D_s^*})^2+ q^2 \Big] \Bigg\} \\
\left(\frac{d \Gamma}{dq^2}\right)_{\rm int}&=&\frac{g(q^2) f(q^2) \tau_{1/2}(q^2)}{27\, m_{D_{s1}^{\prime}}^3m_{D_s^*}}2\sqrt{2}\lambda(m_{D_{s1}^{\prime}}^2,m_{D_s^*}^2,q^2)(2 m_\ell^2+q^2)(m_{D_{s1}^{\prime}}+m_{D_s^*}) \nn \\
&\times&
\Bigg\{g_1^S(m_{D_{s1}^{\prime}}-m_{D_s^*})
+2\frac{g_2^S \, q^2}{\Lambda}\Bigg\}  \,\,\, . 
\eea

\noindent {\fbox{$\bf D_{s1}\to D_s^*  \ell^+  \ell^-$}}
\bea
\left(\frac{d \Gamma}{dq^2}\right)_c&=&g(q^2) [\tau_{3/2}(q^2)]^2
\frac{\lambda^2(m_{D_{s1}}^2,m_{D_s^*}^2,q^2)}{162 \, m_{D_{s1}}^5\,m_{D_s^*}^3}
[(m_{D_{s1}}+m_{D_s}^*)^2+5q^2](2m_\ell^2+q^2)  \\
\left(\frac{d \Gamma}{dq^2}\right)_s&=&g(q^2) f^2(q^2) \frac{h_T^2 q^2}{\Lambda^4\,324 \, m_{D_{s1}}^5m_{D_s^*}}
[5(m_{D_{s1}}+m_{D_s^*})^2+q^2]
\lambda^2(m_{D_{s1}}^2,m_{D_s^*}^2,q^2) (2m_\ell^2+q^2) \nn \\ \\
\left(\frac{d \Gamma}{dq^2}\right)_{\rm int}&=&g(q^2) f(q^2) \tau_{3/2}(q^2)\frac{\sqrt{2}h_T q^2}{\Lambda^2 \,81\, m_{D_{s1}}^5m_{D_s^*}^2}(m_{D_{s1}}+m_{D_s^*})
\lambda^2(m_{D_{s1}}^2,m_{D_s^*}^2,q^2) (2m_\ell^2+q^2) \,\,\, .  \nn \\
\eea
%


\noindent {\fbox{$\bf D_{s2}^*\to D_s^*  \ell^+  \ell^-$}}
\bea
\left(\frac{d \Gamma}{dq^2}\right)_c&=&g(q^2) [\tau_{3/2}(q^2)]^2
\frac{\lambda^2(m_{D_{s2}^*}^2,m_{D_s^*}^2,q^2)}{270 \, m_{D_{s2}^*}^5\,m_{D_s^*}^3}
[5(m_{D_{s2}^*}+m_{D_s}^*)^2+7q^2](2m_\ell^2+q^2)  \\
\left(\frac{d \Gamma}{dq^2}\right)_s&=&g(q^2) f^2(q^2) \frac{h_T^2 q^2}{\Lambda^4\,540 \, m_{D_{s2}^*}^5m_{D_s^*}}
[7(m_{D_{s2}^*}+m_{D_s^*})^2+5q^2]
\lambda^2(m_{D_{s2}^*}^2,m_{D_s^*}^2,q^2) (2m_\ell^2+q^2) \nn \\ \\
\left(\frac{d \Gamma}{dq^2}\right)_{\rm int}&=&-g(q^2) f(q^2) \tau_{3/2}(q^2)\frac{2\sqrt{2}h_T q^2}{\Lambda^2 \,45 \, m_{D_{s2}^*}^5m_{D_s^*}^2}(m_{D_{s2}^*}+m_{D_s^*})
\lambda^2(m_{D_{s2}^*}^2,m_{D_s^*}^2,q^2) (2m_\ell^2+q^2) \nn . \\
\eea
\section{Numerics}\label{sec:numerics}
The expressions of the decay distributions involve  the universal form factors $\tau_{1/2}(w)$ and $\tau_{3/2}(w)$, and  the low-energy couplings in \eqref{LHSV1}, \eqref{LHSV2} and \eqref{LHTV2}. Numerical results can be given using determinations of such  quantities available in the literature.  

For the form factors $\tau_{1/2}$ and $\tau_{3/2}$, the  parametrization 
\be
\tau_i(w)=\tau_i(1)[1-(w-1) \rho_i^2] \label{taui}
\ee
allows us to encode the uncertainties in the value at the zero-recoil point $w=1$ and in the slope. We use the values  in Table \ref{tab:input},  obtained from data on semileptonic $B$ decays to positive-parity charmed mesons \cite{Bernlochner:2017jxt}, which for  $\tau_{1/2}$  agree with the computation in \cite{Colangelo:1998ga}. The value of $g_1^S$ in Table \ref{tab:input} is an estimate obtained  using $D \to K^*$ semileptonic form factors \cite{Casalbuoni:1996pg}. The value of $g_2^S$ comes from a light-cone QCD sum rule  computation of the decay amplitude of the positive-parity mesons to real photons \cite{Colangelo:2005hv}, together with VMD. The value of  $h_T$  is obtained from the  strong decay widths of the excited charmed mesons \cite{Campanella:2018xev}. 

 The relative phases of the two amplitudes describing the photon coupled to $c$ and $\bar s$ depend on the relative phases of the  functions $\tau_i$ and the strong couplings.
  Since such phases are not fixed, we consider the extreme cases: case A where the product of the functions $\tau_i$  and the strong couplings is positive (a relative phase between $g_1^S$ and $g_2^S$ is ignored), and  case B where the product is negative.   
In the experimental analyses the hadronic quantities are parameters to be determined from data.
\begin{table}[!tb]
\center{\begin{tabular}{|l|l|}
\hline
$\alpha(M_Z) = 1/127.9$\hfill\cite{ParticleDataGroup:2022pth}& 	$m_{D_s} =1968.35 \pm 0.07 \, {\rm MeV}$\hfill\cite{ParticleDataGroup:2022pth}\\
$m_\mu=105.66 \, {\rm MeV}$\hfill\cite{ParticleDataGroup:2022pth}	
& $m_{D_s^*} = 2112.2 \pm 0.4 \, {\rm MeV}$\hfill\cite{ParticleDataGroup:2022pth}\\
& $m_{D_{s0}^*} = 2317.8 \pm 0.5\, {\rm MeV}$\hfill\cite{ParticleDataGroup:2022pth}\\
\cline{1-1}
$g_V=5.8$  \hfill \cite{Casalbuoni:1996pg}& $m_{D_{s1}^\prime} = 2459.5 \pm 0.6 \, {\rm MeV}$\hfill\cite{ParticleDataGroup:2022pth}\\
$\Lambda=1$ GeV  \hfill \cite{Casalbuoni:1996pg}& $m_{D_{s1}} = 2535.11 \pm 0.06  \, {\rm MeV}$ \quad \hfill\cite{ParticleDataGroup:2022pth}\\
$m_\phi=1019.461 \pm 0.016 \, {\rm MeV}$\hfill\cite{ParticleDataGroup:2022pth}		
& $\Gamma({D_{s1})} = 0.92 \pm 0.05 \, {\rm MeV}$\hfill\cite{ParticleDataGroup:2022pth}\\
$\Gamma_\phi=4.249 \pm 0.013 \, {\rm MeV}$\hfill\cite{ParticleDataGroup:2022pth}&  $m_{D_{s2}^*} = 2569.1 \pm 0.8 \, {\rm MeV}$\hfill\cite{ParticleDataGroup:2022pth}\\
${\cal B} (\phi \to \mu^+ \mu^-)=(2.85 \pm 0.19) \times 10^{-4}$ \quad \hfill\cite{ParticleDataGroup:2022pth} &
$\Gamma(D_{s2}^*) = 16.9 \pm 0.7 \, {\rm MeV}$\hfill\cite{ParticleDataGroup:2022pth} \\
\cline{2-2}
$f_\phi=208 \pm 7$ MeV \hfill\cite{ParticleDataGroup:2022pth}&$\tau_{1/2}(1)=0.70\pm 0.21$ \hfill\cite{Bernlochner:2017jxt} \\
\cline{1-1}
$|g_1^S|=0.10$  \hfill \cite{Casalbuoni:1996pg} &$\hat \rho_{1/2}^2=-0.2\pm 1.4$ \hfill\cite{Bernlochner:2017jxt} \\
$|g_2^S|=0.29 \pm 0.03$  \hfill \cite{Colangelo:2005hv} & $\tau_{3/2}(1)=0.70\pm 0.07$ \quad \hfill\cite{Bernlochner:2017jxt}  \\
$|h_T|=0.23 \pm 0.09$  \hfill\cite{Campanella:2018xev}&$\hat \rho_{3/2}^2=1.6 \pm 0.2$ \hfill\cite{Bernlochner:2017jxt}   \\
\hline
\end{tabular}  }
\caption {Input parameters in the numerical analysis.}
\label{tab:input}
\end{table}

\begin{table}[!b]
\center{\begin{tabular}{| l | c | c |c|c|}
\hline
 & \small{Width $\times10^{8}$ (GeV)}& \small{Width $\times10^{8}$ (GeV)} & \small{BR $\times10^{6}$  }& \small{BR $\times10^{6}$ } \\
 & \small{(case A) }& \small{(case B)}& \small{(case A) }& \small{(case B)}\\
\hline
$D_{s1}^\prime \to D_s    \mu^+ \mu^-$&$ 4.7\pm 0.8$ & $8.1\pm 1.3  $ &&\\
$D_{s1}^\prime \to D_s^* \mu^+ \mu^-$&$ 2.1\pm 0.3 $ &$ 2.7\pm 0.4 $ &&\\
$D_{s1} \to D_s \mu^+ \mu^-$              &$ 2.9\pm 0.6 $ &  $2.8\pm 0.6 $ &$ 31.4\pm 6.4 $ &  $30.7\pm 6.3 $\\
$D_{s1} \to D_s^* \mu^+ \mu^-$            &$0.18\pm 0.04$ &  $0.19\pm 0.04$ &$1.95\pm 0.4$ &  $2.1\pm 0.4$\\
$D_{s2}^* \to D_s \mu^+ \mu^-$            &$0.05\pm 0.06$ &  $0.22\pm 0.13 $ &  $0.03\pm 0.04 $    &$0.13\pm 0.08$\\
$D_{s2}^* \to D_s^* \mu^+ \mu^-$         &$0.96\pm 0.19$ &  $0.87\pm 0.17$ &$0.6\pm 0.1$ &  $0.5\pm 0.1$\\
\hline
\end{tabular}  }
\caption {Decay widths  of Dalitz modes for the  two cases of maximal interference. Branching fractions are displayed only for $D_{s1}$ and $D_{s2}^*$ for which the full widths are measured.}
\label{tab:res}
\end{table}

Table \ref{tab:res}  contains the decay widths  corresponding to the maximal interference cases. We also display the branching fractions for $D_{s1}$ and $D^*_{s2}$, for which the full widths are measured. The effect of the interference is large in $D_{s1}^\prime \to D_s    \mu^+ \mu^-$ and  
$D_{s2}^* \to D_s \mu^+ \mu^-$, for the other modes the interference contribution is small.
\begin{figure}[!t]
\begin{center}
\includegraphics[width = 0.40\textwidth]{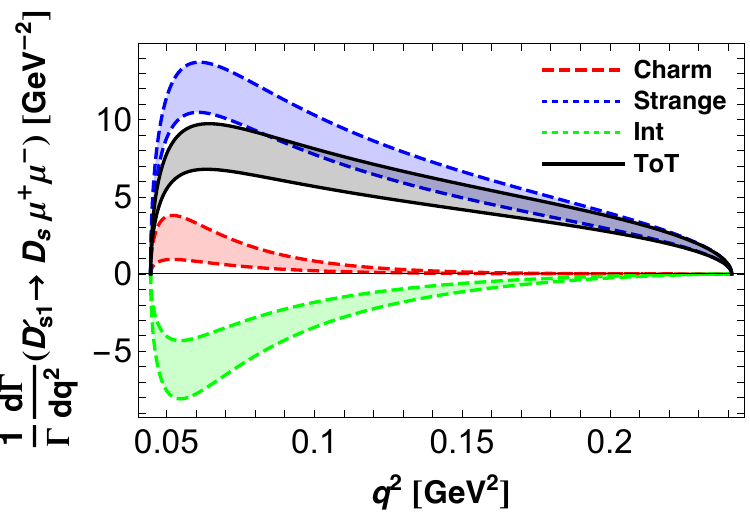}
\includegraphics[width = 0.40\textwidth]{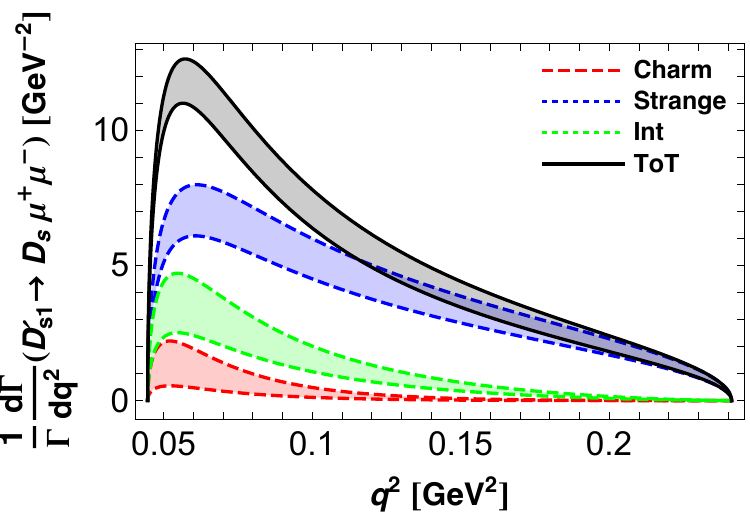} \\
\includegraphics[width = 0.40\textwidth]{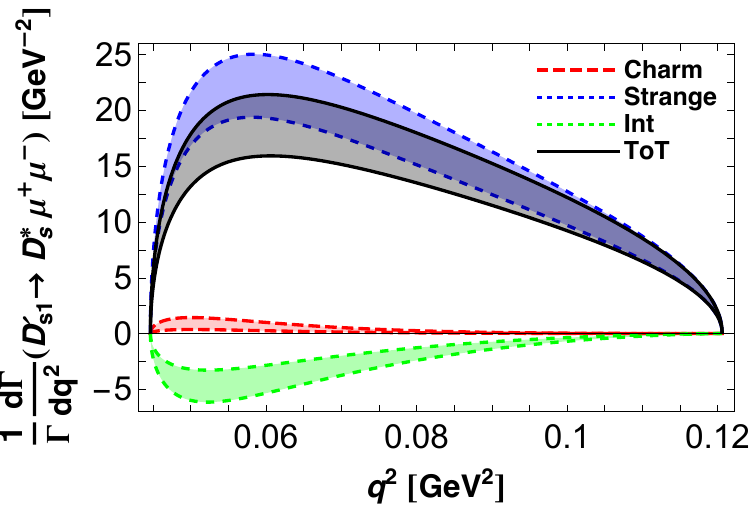}
\includegraphics[width = 0.40\textwidth]{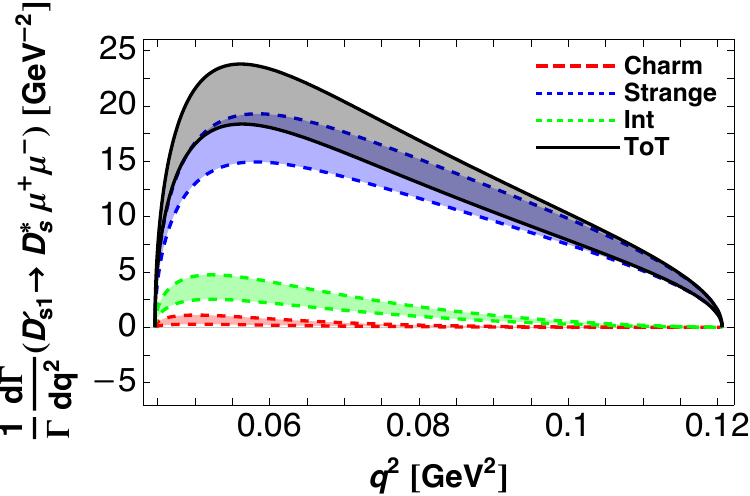}
\caption{\small  Distribution $\dd \frac{1}{\Gamma} \frac{d \Gamma}{d q^2}$   for the modes  $D_{s1}^\prime \to D_s \mu^+ \mu^-$ (top) and   $D_{s1}^\prime  \to D^*_s \mu^+ \mu^-$ (bottom panels) for the two signs of the interference between the amplitudes with the photon coupled to the charm and $\bar s$  quark (case A, left panels;  case B, right panels). The distributions for the photon coupled to quarks and the interference term are separately shown, together with their sum.}\label{fig:Dp1Ds}
\end{center}
\end{figure}
Such effects can be better observed in the decay distributions.
The  dilepton invariant mass distributions for the modes $D_{s1}^\prime \to D_s  \mu^+  \mu^-$ and $D_{s1}^\prime \to D_s^*  \mu^+  \mu^-$ is depicted in Fig.~\ref{fig:Dp1Ds} 
for the   maximal interferences. The interference term is sizable in the former mode.
 The amplitude corresponding to the photon coupled to the light quark gives the largest contribution. The distributions for $D_{s1} \to D_s^{(*)}  \mu^+  \mu^-$ and  $D_{s2}^* \to D_s^{(*)}  \mu^+  \mu^-$ are  shown  in figs. \ref{fig:D1Ds} and \ref{fig:D2Ds}, with a sizable interference term  visible in $D_{s2}^* \to D_s  \mu^+  \mu^-$.
%
\begin{figure}[t]
\begin{center}
\includegraphics[width = 0.40\textwidth]{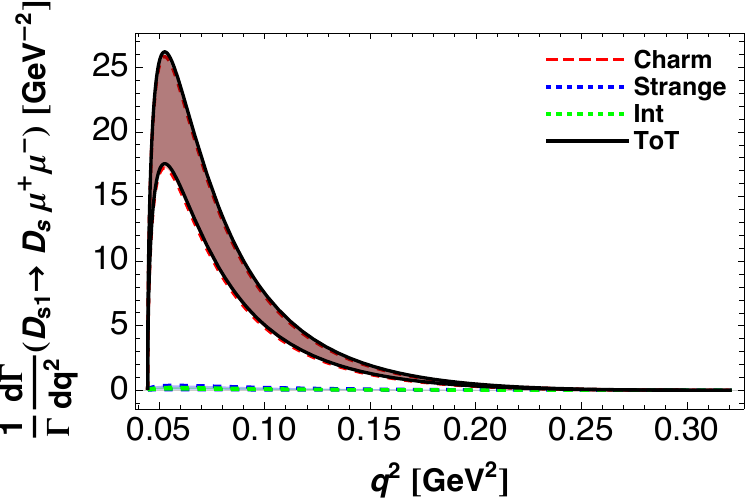}
\includegraphics[width = 0.40\textwidth]{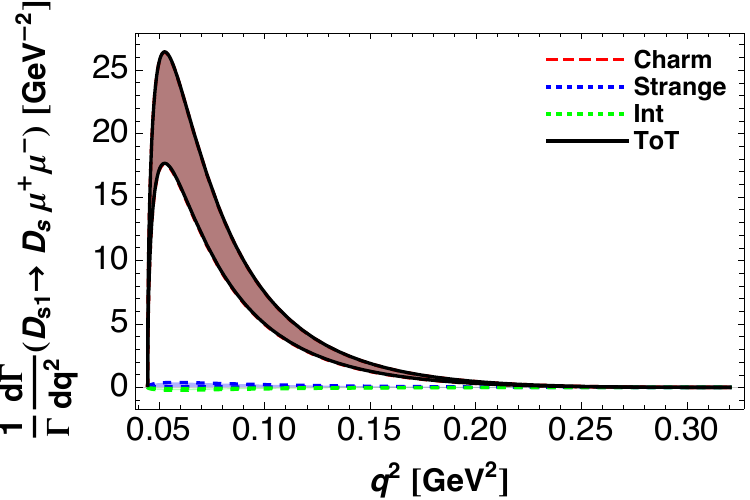}\\
\includegraphics[width = 0.40\textwidth]{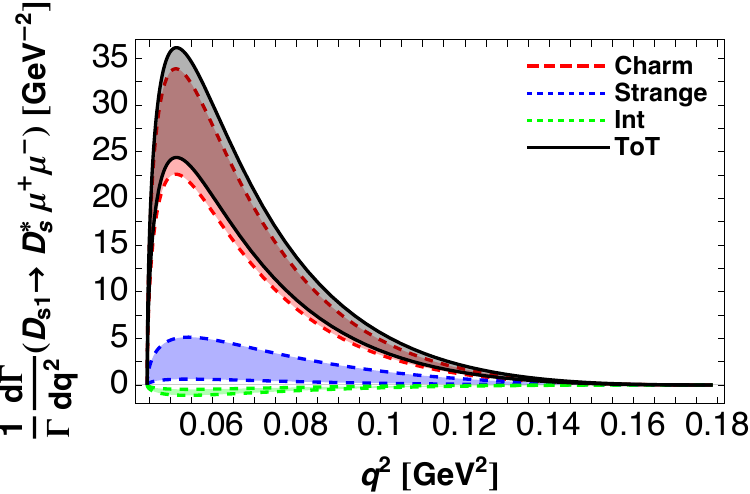}
\includegraphics[width = 0.40\textwidth]{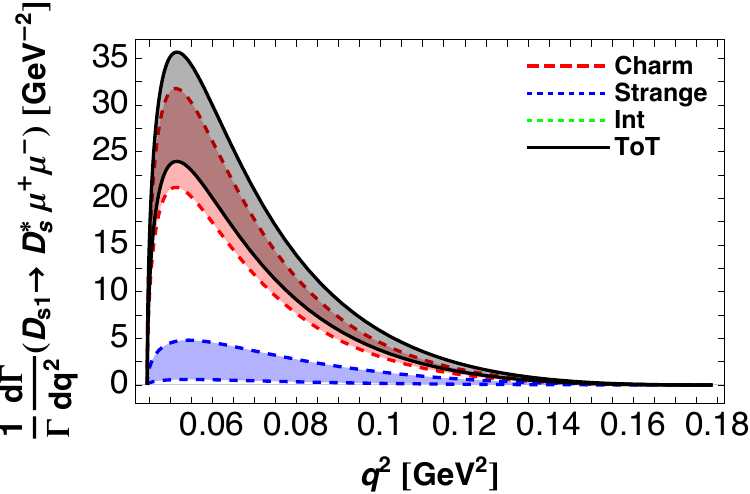}
\caption{\small  Distribution  $\dd \frac{1}{\Gamma} \frac{d \Gamma}{d q^2}$  for  $D_{s1} \to D_s \mu^+ \mu^-$ (top) and  $D_{s1} \to D_s^* \mu^+ \mu^-$ (bottom panels). Notations as in Fig.~\ref{fig:Dp1Ds}.  }\label{fig:D1Ds}
\end{center}
\end{figure}
\begin{figure}[!t]
\begin{center}
\includegraphics[width = 0.40\textwidth]{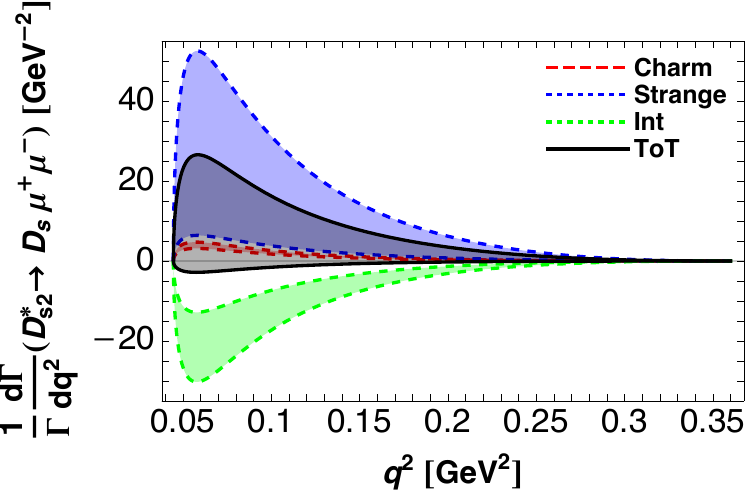}
\includegraphics[width = 0.40\textwidth]{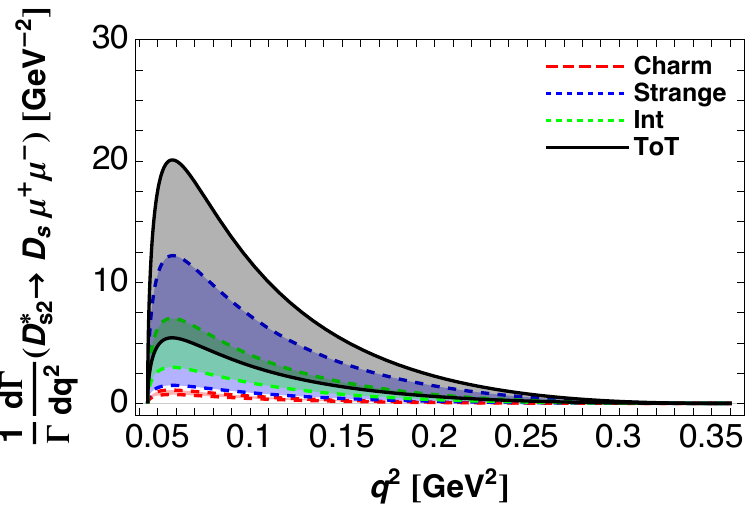}\\
\includegraphics[width = 0.40\textwidth]{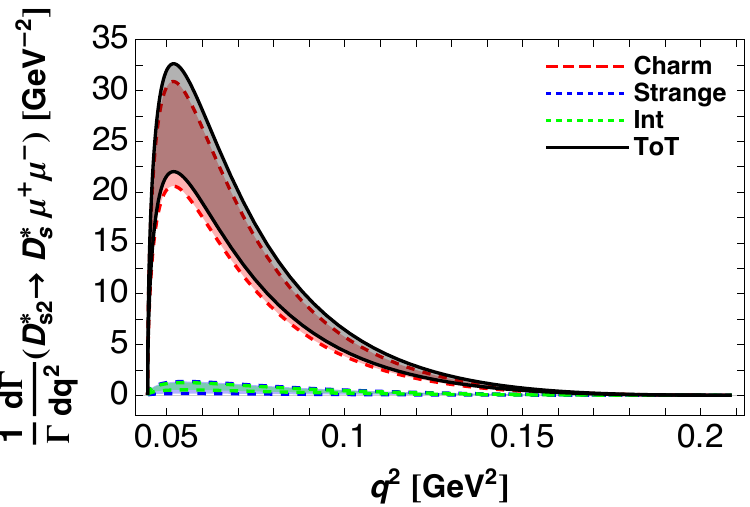}
\includegraphics[width = 0.40\textwidth]{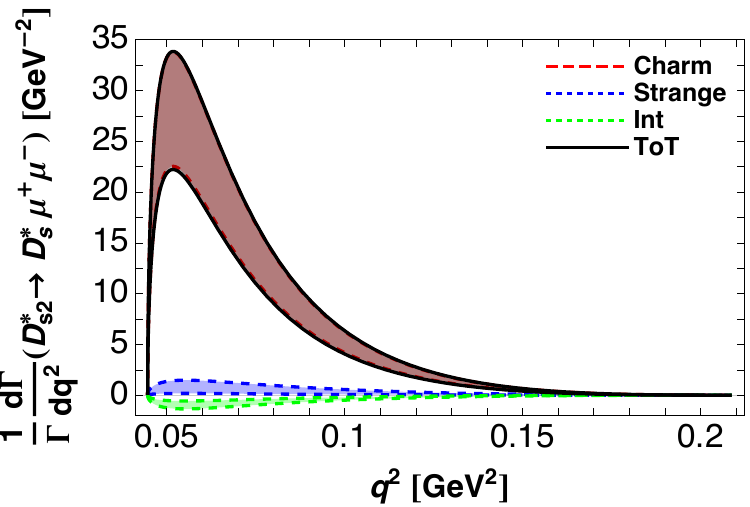}
\caption{\small  Distribution $\dd \frac{1}{\Gamma} \frac{d \Gamma}{d q^2}$   for  $D_{s2}^* \to D_s \mu^+ \mu^-$ (top) and $D_{s2}^* \to D^*_s \mu^+ \mu^-$ (bottom panels).  Notations  as in Fig.~\ref{fig:Dp1Ds}. }\label{fig:D2Ds}
\end{center}
\end{figure}

\begin{figure}[!t]
\begin{center}
\includegraphics[width = 0.31\textwidth]{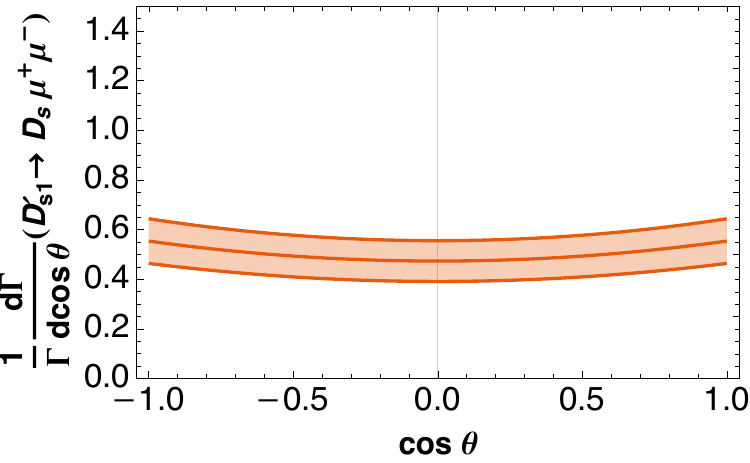} 
\includegraphics[width = 0.31\textwidth]{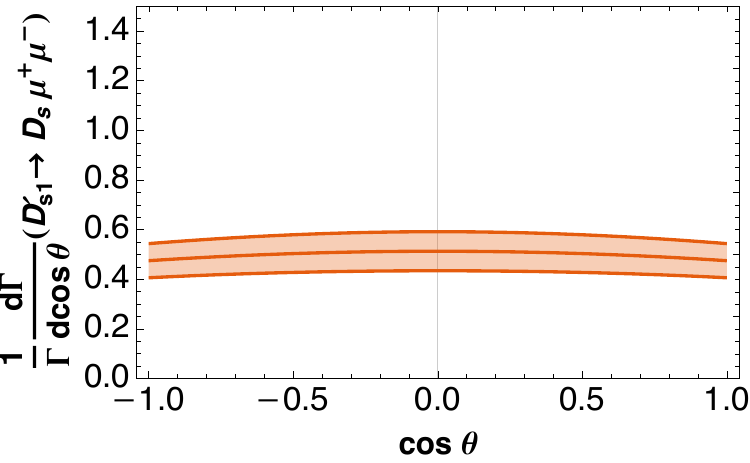} \\
\includegraphics[width = 0.31\textwidth]{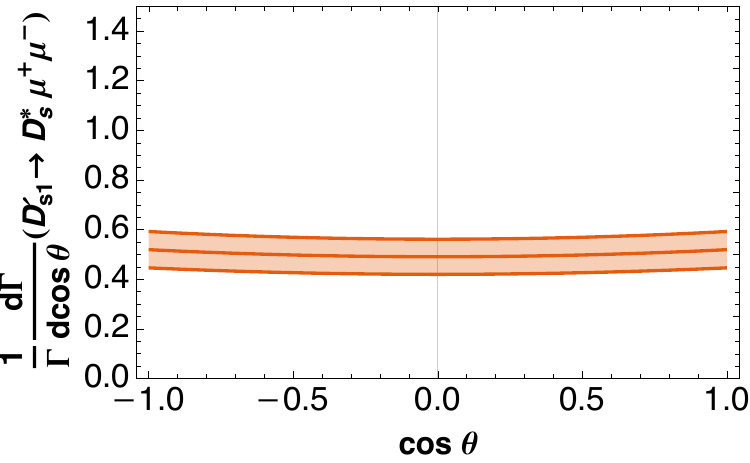}
\includegraphics[width = 0.31\textwidth]{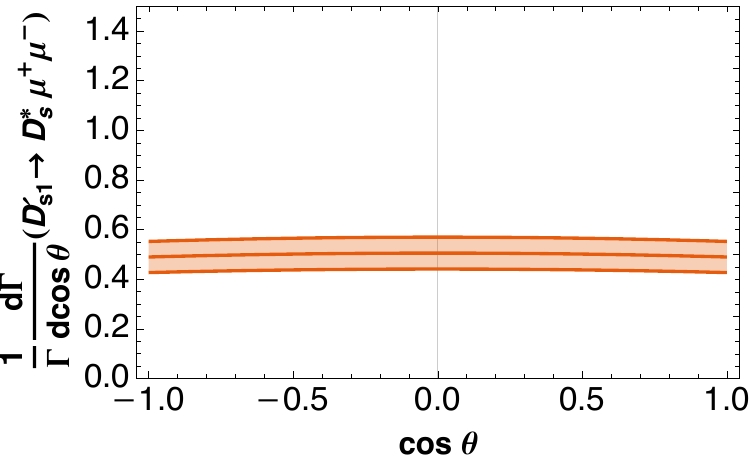}\\
\includegraphics[width = 0.31\textwidth]{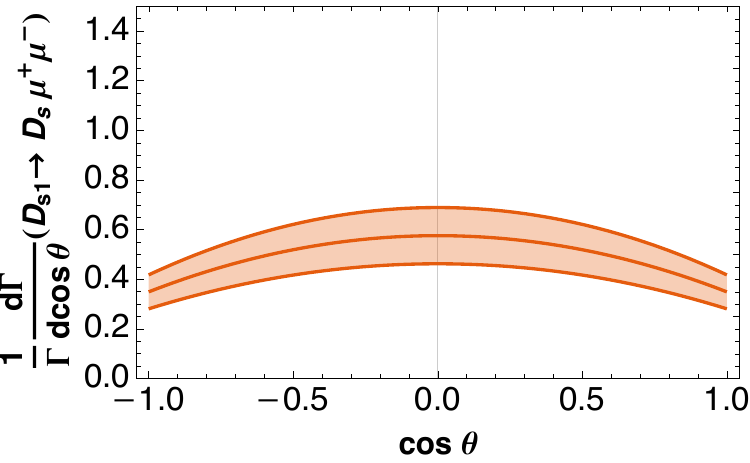}
\includegraphics[width = 0.31\textwidth]{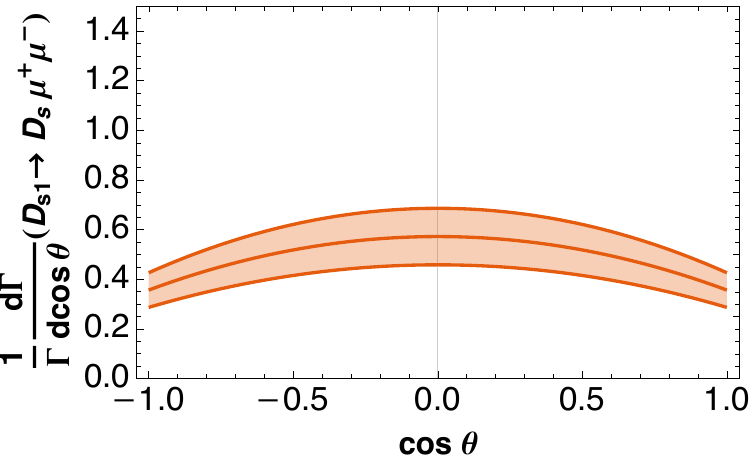}\\
\includegraphics[width = 0.31\textwidth]{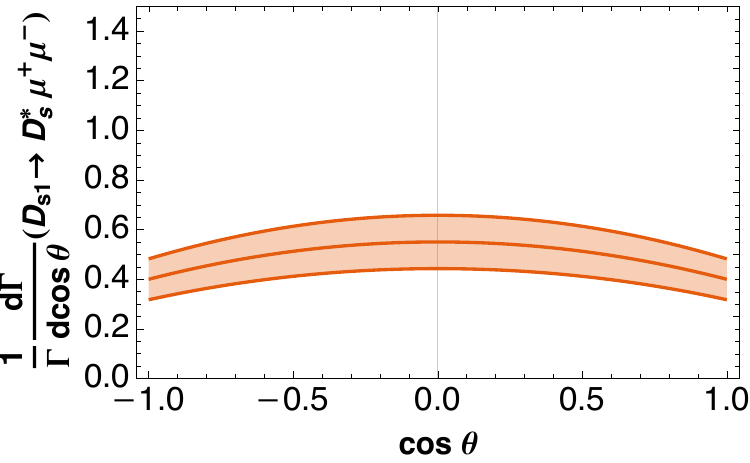}
\includegraphics[width = 0.31\textwidth]{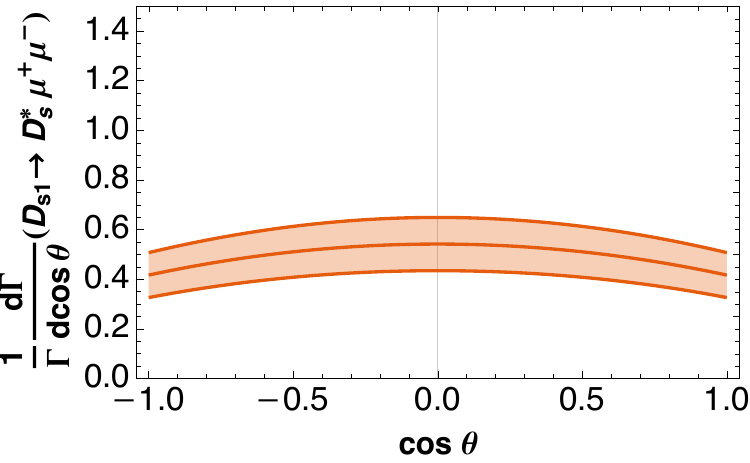}\\
\includegraphics[width = 0.31\textwidth]{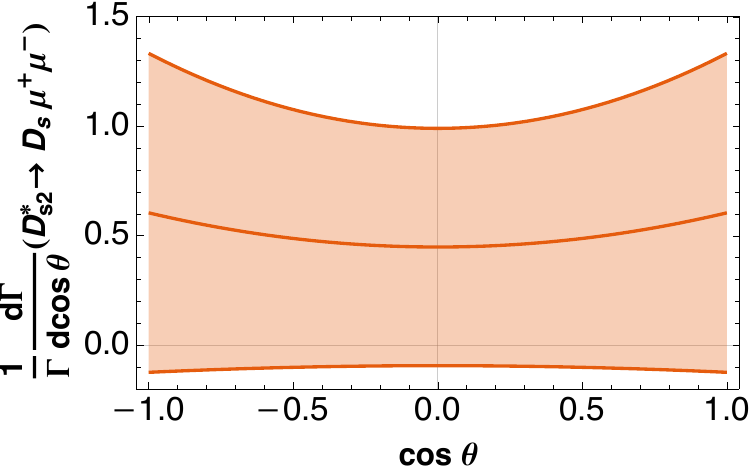}
\includegraphics[width = 0.31\textwidth]{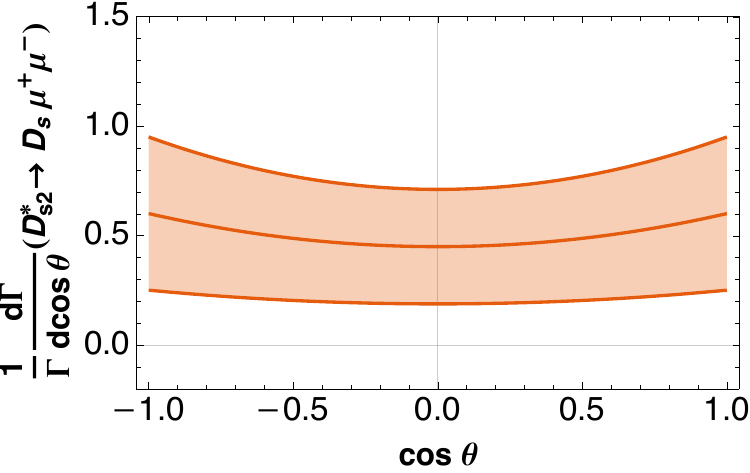}\\
\includegraphics[width = 0.31\textwidth]{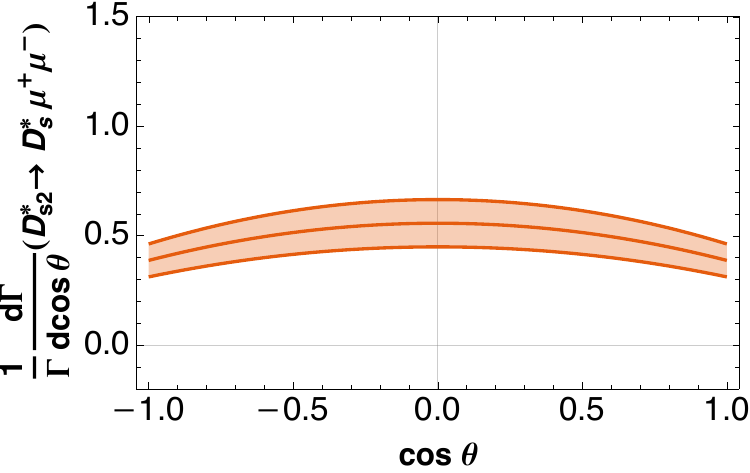}
\includegraphics[width = 0.31\textwidth]{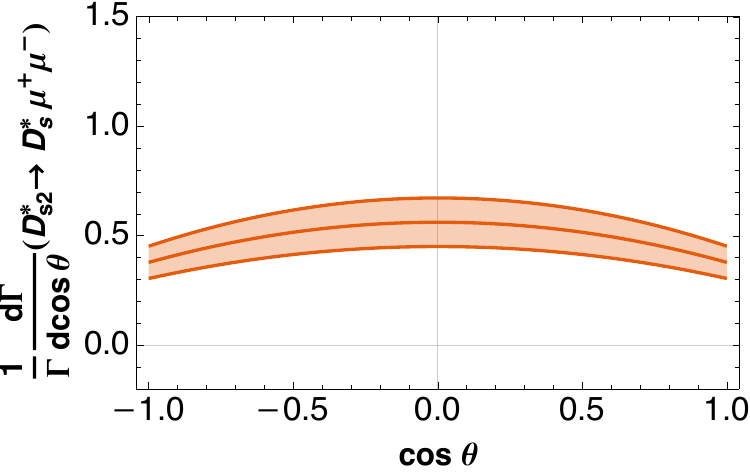}
\caption{\small  Angular distribution $\dd \frac{1}{\Gamma} \frac{d \Gamma}{d \cos \theta}$ for  the processes (from top to bottom)
$D_{s1}^\prime \to D_s \mu^+ \mu^-$, 
$D_{s1}^\prime \to D_s^* \mu^+ \mu^-$, 
$D_{s1} \to D_s \mu^+ \mu^-$, 
$D_{s1} \to D_s^* \mu^+ \mu^-$, 
$D_{s2}^* \to D_s \mu^+ \mu^-$, 
$D_{s2}^* \to D_s^* \mu^+ \mu^-$.
 Left panels  correspond to case A, right panels to case B for the maximal interference. }\label{fig:DptoDang}
\end{center}
\end{figure}

The angular distributions are displayed in Fig.~\ref{fig:DptoDang} in the cases of extremal interferences. From the plots one finds that the $D_{s2}^* \to D_s \mu^+ \mu^-$ mode  is affected by the largest uncertainty.

\begin{figure}[!b]
\begin{center}
\includegraphics[width = 0.4\textwidth]{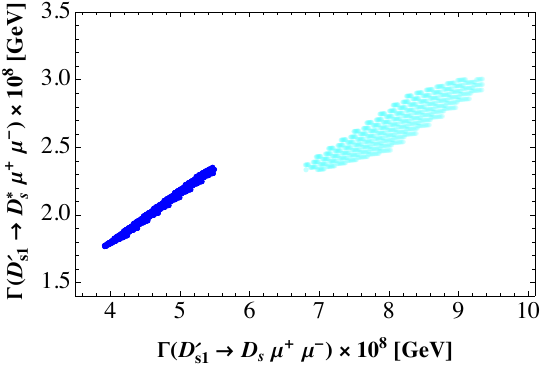} \hspace*{0.2cm}
\includegraphics[width =  0.4\textwidth]{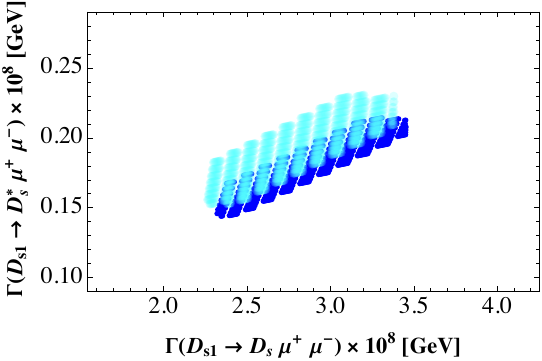}\\ \vspace*{0.4cm}
\includegraphics[width =  0.4\textwidth]{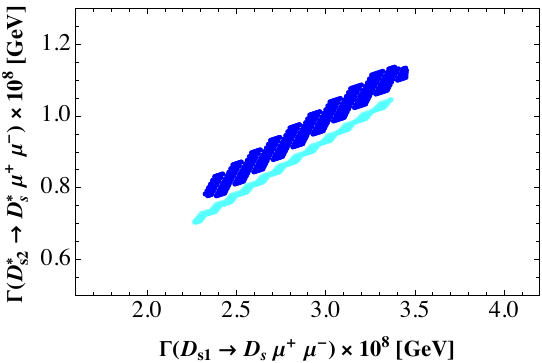} \hspace*{0.2cm}
\includegraphics[width =  0.4\textwidth]{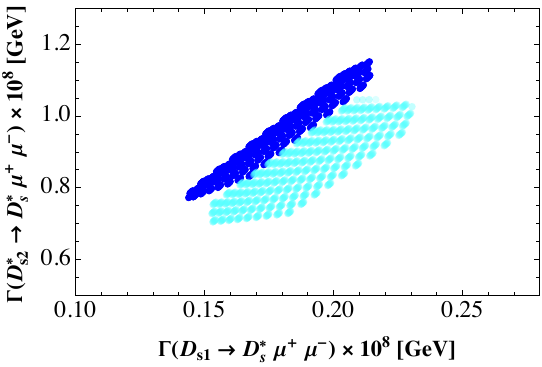} 
\caption{\small  Correlations between decay widths for different Dalitz decays. The dark and light regions correspond to the maximal interference cases A and B, respectively. }\label{fig:correlations}
\end{center}
\end{figure}

Since the amplitudes of different modes  involve  the same hadronic parameters, the decay rates and other observables are correlated, as shown 
in Fig.~\ref{fig:correlations} for the decay widths. Varying the parameters of the ranges  in Table~\ref{tab:input},
a positive correlation is found between  
$\Gamma(D_{s1}^\prime \to D_s    \mu^+ \mu^-)$  and   $\Gamma(D_{s1}^\prime \to D_s^*    \mu^+ \mu^-)$  for both cases of interference,
as well as for the widths of processes with  $\dd s_\ell^P=\frac{3}{2}^+$ particles.
The experimental confirmation of such a  behaviour would further  support  the classification of $D_{sJ}^{(*)}$ in heavy quark spin doublets.

The requests for the experimental analyses are the optimized reconstruction of $D_s$ and $D_s^*$ and of their momentum,  and the precise measurement of the lepton momenta to access $q^2$ and angular distributions. Both the requirements are well satisfied by a hadron facility as the CERN  LHC, in particular by the LHCb experiment which can  exploit the large charm production rate needed for signals with branching fractions as in Table \ref{tab:res}. Measurements are also feasible at a lepton facility such as the Belle II experiment at KEK, in particular when all the statistics are available. 

 In the experimental analyses, the expressions for the amplitudes and the distributions can be parametrically used, leaving the strong couplings $g_{1,2}^S$ and $h_T$ and the parameters of the universal functions $\tau_i$ in Table \ref{tab:input} as quantities to be measured. This will allow an interesting comparison with the determinations from different observables (strong decay widths of positive-parity charmed mesons and semileptonic widths of $B_{q,s}$ decays to $D_{q,s J}^{(*)}$). 

From a general viewpoint, we emphasize that the expressions given in the previous sections and used in the examples presented here have been worked out under the assumption that the four lightest $D_{sJ}^{(*)}$ belong to heavy quark spin doublet. In such a case, the formulas involve the heavy quark effective theory of QCD together with the VMD model to describe  the coupling of the virtual photon to the strange quark. All such features can be probed using the wealth of observables that can be constructed: for single modes,  with rates and distributions in dilepton mass and angles, and comparing different modes, in pairs or all together. This justifies the great interest for the processes we have discussed. 

\section{Conclusions}

We have presented an analysis of the Dalitz decays of  the positive-parity $D_{sJ}^{(*)}$ charmed mesons, $D_{sJ}^{(*)} \to D_s^{(*)} \ell^+  \ell^-$, with $J=0,1,2$ and $\ell=e, \mu$.
The study is based on the classification of the heavy mesons in spin doublets. The amplitudes are expressed in terms of universal form factors and of effective strong couplings, and can be computed using information from different processes. We have discussed how correlations can be established among different observables: Their experimental confirmation would  further  support our  classification scheme for the $D_{sJ}^{(*)}$ charmed mesons.

\section*{Acknowledgements}
\noindent
We thank Francesco Debernardis and Marco Pappagallo for discussions.
The Feynman diagrams have been generated using the TikZ-Feynman package \cite{Ellis:2016jkw}.
The  research has been carried out within the INFN projects  (Iniziative Specifiche)  QFT-HEP and SPIF.
 
\appendix
\section{Functions in the angular distribution Eq.~\eqref{eq:doubled}}
 For the various Dalitz processes,  the expressions of the functions $A_X(q^2)$ and $B_X(q^2)$ in the angular distribution in Eq.~\eqref{eq:doubled} are given below.

\vspace{1cm}
\noindent{\fbox {$ \bf {D_{s1}^\prime \to D_s \ell^+  \ell^-}$}}
\bea
A_c(q^2)&=&g(q^2) \, \big[ \tau_{1/2}(q^2) \big]^2 \, \frac{[ ( m_{D_{s1}^\prime} - m_{D_s} )^2 - q^2 ]^2}{18 \, m_{D_{s1}^\prime}^3 \, m_{D_s}} \, q^2 \, [ ( m_{D_{s1}^\prime} + m_{D_s} )^2 + q^2 + 4 \, m_\ell^2 ] \\
B_c(q^2)& = &- g(q^2) \, \big[ \tau_{1/2}(q^2) \big]^2 \, \frac{[ ( m_{D_{s1}^\prime} - m_{D_s} )^2 - q^2 ]^2}{18 \, m_{D_{s1}^\prime}^3 \, m_{D_s}} \, [ ( m_{D_{s1}^\prime} + m_{D_s} )^2 - q^2 ] \, ( q^2 - 4 \, m_\ell^2 ) \quad
\eea
\bea
A_s(q^2) & = &g(q^2) \, f^2(q^2)  \, \frac{[ ( m_{D_{s1}^\prime} + m_{D_s} )^2 - q^2 ]^2}{144 \, m_{D_{s1}^\prime}^3 \, m_{D_s}} \, q^2 \nn \\
& \times& \bigg\{ \big( g_1^S \big)^2 \, [ ( m_{D_{s1}^\prime} - m_{D_s} )^2 + q^2 + 4 \, m_\ell^2 ] + \frac{8 \, g_1^S \, g_2^S}{\Lambda} \, \big( m_{D_{s1}^\prime} - m_{D_s} \big) \, ( q^2 + 2 \, m_\ell^2 ) + \nn \\
&  + &\frac{4 \, (g_2^S)^2}{\Lambda^2} \, \big[ q^4 + q^2 \, ( m_{D_{s1}^\prime} - m_{D_s} )^2 + 4 \, m_\ell^2 \, ( m_{D_{s1}^\prime} - m_{D_s} )^2 \big] \bigg\} 
\eea
\bea
B_s(q^2)& =& - g(q^2) \,  f^2(q^2)  \, \frac{[ ( m_{D_{s1}^\prime} + m_{D_s} )^2 - q^2 ]^2}{144 \, m_{D_{s1}^\prime}^3 \, m_{D_s}} \, [ ( m_{D_{s1}^\prime} - m_{D_s} )^2 - q^2 ] \, ( q^2 - 4 \, m_\ell^2 ) \nn \\
& \times& \bigg\{ \big( g_1^S \big)^2 - \frac{4 \, (g_2^S)^2}{\Lambda^2} \, q^2 \bigg\} 
\eea
\bea
A_\text{int}(q^2) & = &g(q^2) \, f(q^2) \,  \tau_{1/2}(q^2) \, \frac{q^2 \, \lambda(m_{D_{s1}^\prime}^2, m_{D_s}^2, q^2)}{18 \, \sqrt{2} \, m_{D_{s1}^\prime}^3 \, m_{D_s}} \nn \\
& \times& \bigg\{ g_1^S \, [ m_{D_{s1}^\prime}^2 - m_{D_s}^2 + q^2 + 4 \, m_\ell^2 ] + \frac{4 \, g_2^S}{\Lambda} \, [ q^2 \, m_{D_{s1}^\prime} + 2 \, m_\ell^2 \, ( m_{D_{s1}^\prime} - m_{D_s} ) ] \bigg\} \\
B_\text{int}(q^2) & = &- g(q^2) \, f(q^2) \,  \, \tau_{1/2}(q^2) \, \frac{\lambda(m_{D_{s1}^\prime}^2, m_{D_s}^2, q^2)}{18 \, \sqrt{2} \, m_{D_{s1}^\prime}^3 \, m_{D_s}} \, ( q^2 - 4 \, m_\ell^2 ) \nn \\
& \times& \bigg\{ g_1^S \, [ m_{D_{s1}^\prime}^2 - m_{D_s}^2 - q^2 ] + \frac{4 \, g_2^S}{\Lambda} \, q^2 \, m_{D_s}  \bigg\} \,\, . 
\eea

\vspace{1cm}
\noindent {\fbox{$\bf D_{s1} \to D_s \ell^+  \ell^-$}}
\begin{align}
A_c (q^2)& = g(q^2) \, \big[ \tau_{3/2}(q^2) \big]^2 \, \frac{\lambda^2(m_{D_{s1}}^2, m_{D_s}^2, q^2)}{432 \, m_{D_{s1}}^5 \, m_{D_s}^3} \, q^2 \, [ 4 \, ( m_{D_{s1}} + m_{D_s} )^2 + q^2 + 4 \, m_\ell^2 ] \\
B_c(q^2)& = - g(q^2) \, \big[ \tau_{3/2}(q^2) \big]^2 \, \frac{\lambda^2(m_{D_{s1}}^2, m_{D_s}^2, q^2)}{432 \, m_{D_{s1}}^5 \, m_{D_s}^3} \, ( q^2 - 4 \, m_\ell^2 ) \, [ 4 \, ( m_{D_{s1}} + m_{D_s} )^2 - q^2 ] \\
A_s (q^2)& = g(q^2) \,  f^2(q^2)  \, \frac{h_T^2}{\Lambda^4} \, \frac{\lambda^2(m_{D_{s1}}^2, m_{D_s}^2, q^2)}{864 \, m_{D_{s1}}^5 \, m_{D_s}} \, q^2 \, [ 4 \, q^4 + ( q^2 + 4 \, m_\ell^2 ) \, ( m_{D_{s1}} + m_{D_s} )^2 ] \\
B_s (q^2)& = g(q^2) \,  f^2(q^2)  \, \frac{h_T^2}{\Lambda^4} \, \frac{\lambda^2(m_{D_{s1}}^2, m_{D_s}^2, q^2)}{864 \, m_{D_{s1}}^5 \, m_{D_s}} \, q^2 \, ( q^2 - 4 \, m_\ell^2 ) \, [ ( m_{D_{s1}} + m_{D_s} )^2 - 4 \, q^2 ] \\
A_\text{int} (q^2) & = - g(q^2) \, f(q^2) \, \tau_{3/2}(q^2) \, \frac{h_T}{\Lambda^2} \, \frac{\lambda^2(m_{D_{s1}}^2, m_{D_s}^2, q^2)}{216 \, \sqrt{2} \, m_{D_{s1}}^5 \, m_{D_s}^2} \, q^2 \, ( 3 \, q^2 - 4 \, m_\ell^2 ) \, ( m_{D_{s1}} + m_{D_s} ) \\
B_\text{int}(q^2) & = g(q^2) \, f(q^2) \, \tau_{3/2}(q^2) \, \frac{h_T}{\Lambda^2} \, \frac{5 \, \lambda^2(m_{D_{s1}}^2, m_{D_s}^2, q^2)}{216 \, \sqrt{2} \, m_{D_{s1}}^5 \, m_{D_s}^2} \, q^2 \, ( q^2 - 4 \, m_\ell^2 ) \, ( m_{D_{s1}} + m_{D_s} )
\end{align}

\vspace{1cm}
\noindent{ \fbox{$ \bf D_{s2}^* \to D_s \ell^+  \ell^-$}}
\bea
A_c (q^2)& = & g(q^2) \, \big[ \tau_{3/2}(q^2) \big]^2 \, \frac{ \lambda^2(m_{D_{s2}^*}^2, m_{D_s}^2, q^2) }{240 \, m_{D_{s2}^*}^5 \, m_{D_s}^3} \, q^2 \, ( q^2 + 4 \, m_\ell^2 ) \\
B_c (q^2)& = & g(q^2) \, \big[ \tau_{3/2}(q^2) \big]^2 \, \frac{ \lambda^2(m_{D_{s2}^*}^2, m_{D_s}^2, q^2) }{240 \, m_{D_{s2}^*}^5 \, m_{D_s}^3} \, q^2 \, ( q^2 - 4 \, m_\ell^2 ) \\
A_s (q^2)& = & g(q^2) \,  f^2(q^2)  \, \frac{h_T^2}{\Lambda^4} \, \frac{ \lambda^2(m_{D_{s2}^*}^2, m_{D_s}^2, q^2) }{480 \, m_{D_{s2}^*}^5 \, m_{D_s}} \, q^2 \, ( q^2 + 4 \, m_\ell^2 ) \, ( m_{D_{s2}^*} + m_{D_s} )^2 \\
B_s (q^2)& = & g(q^2) \,  f^2(q^2)  \, \frac{h_T^2}{\Lambda^4} \, \frac{ \lambda^2(m_{D_{s2}^*}^2, m_{D_s}^2, q^2) }{480 \, m_{D_{s2}^*}^5 \, m_{D_s}} \, q^2 \, ( q^2 - 4 \, m_\ell^2 ) \, ( m_{D_{s2}^*} + m_{D_s} )^2 \\
A_\text{int}(q^2) & = & g(q^2) \, f(q^2) \, \tau_{3/2}(q^2) \, \frac{h_T}{\Lambda^2} \, \frac{\lambda^2(m_{D_{s2}^*}^2, m_{D_s}^2, q^2) }{120 \, \sqrt{2} \, m_{D_{s2}^*}^5 \, m_{D_s}^2} \, q^2 \, ( q^2 + 4 \, m_\ell^2 ) \, ( m_{D_{s2}^*} + m_{D_s} ) \\
B_\text{int} (q^2)& = & g(q^2) \, f(q^2) \, \tau_{3/2}(q^2) \, \frac{h_T}{\Lambda^2} \, \frac{ \lambda^2(m_{D_{s2}^*}^2, m_{D_s}^2, q^2) }{120 \, \sqrt{2} \, m_{D_{s2}^*}^5 \, m_{D_s}^2} \, q^2 \, ( q^2 - 4 \, m_\ell^2 ) \, ( m_{D_{s2}^*} + m_{D_s} ) \,\, . \quad
\eea

\vspace{1cm}
\newpage
\noindent {\fbox{$\bf D_{s0}^*\to D_s^* \, \ell^+ \,  \ell^-$}} \\

The process is kinematically allowed only for $\ell=e$. The functions in \eqref{eq:doubled} are given by:
\bea
A_c (q^2) & = & g(q^2) \, \big[ \tau_{1/2}(q^2) \big]^2 \, \frac{[ ( m_{D_{s0}^*} - m_{D_s^*} )^2 - q^2 ]^2}{6 \, m_{D_{s0}^*}^3 \, m_{D_s^*}} \, q^2 \, [ ( m_{D_{s0}^*} + m_{D_s^*} )^2 + q^2 + 4 \, m_\ell^2 ] \\
B_c (q^2)& =  &- g(q^2) \, \big[ \tau_{1/2}(q^2) \big]^2 \, \frac{[ ( m_{D_{s0}^*} - m_{D_s^*} )^2 - q^2 ]^2}{6 \, m_{D_{s0}^\prime}^3 \, m_{D_s^*}} \, ( q^2 - 4 \, m_\ell^2 ) \, [ ( m_{D_{s0}^*} + m_{D_s^*} )^2 - q^2 ] \qquad \\
A_s (q^2)& = & g(q^2) \,  f^2(q^2)  \, \frac{[ ( m_{D_{s0}^*} + m_{D_s^*} )^2 - q^2 ]^2}{48 \, m_{D_{s0}^*}^3 \, m_{D_s^*}} \, q^2 \nn \\
& \times &  \bigg\{ \big( g_1^S \big)^2 \, [ ( m_{D_{s0}^*} - m_{D_s^*} )^2 + q^2 + 4 \, m_\ell^2 ] + \frac{8 \, g_1^S \, g_2^S}{\Lambda} \, \big( m_{D_{s0}^*} - m_{D_s^*} \big) \, ( q^2 + 2 \, m_\ell^2 )  \nn \\
&  + & \frac{4 \, (g_2^S)^2}{\Lambda^2} \, \big[ q^4 + q^2 \, ( m_{D_{s0}^*} - m_{D_s^*} )^2 + 4 \, m_\ell^2 \, ( m_{D_{s0}^*} - m_{D_s^*} )^2 \big] \bigg\} \\
B_s (q^2)& = & - g(q^2) \,  f^2(q^2)  \, \frac{[ ( m_{D_{s0}^*} + m_{D_s^*} )^2 - q^2 ]^2}{48 \, m_{D_{s0}^*}^3 \, m_{D_s^*}} \, [ ( m_{D_{s0}^*} - m_{D_s^*} )^2 - q^2 ] \, ( q^2 - 4 \, m_\ell^2 ) \nn \\
&  \times & \bigg\{ \big( g_1^S \big)^2 - \frac{4 \, (g_2^S)^2}{\Lambda^2} \, q^2 \bigg\} \\
A_\text{int} (q^2)& = & g(q^2) \, f(q^2) \, \tau_{1/2}(q^2) \, \frac{\lambda(m_{D_{s0}^*}^2, m_{D_s^*}^2, q^2)}{6 \, \sqrt{2} \, m_{D_{s0}^*}^3 \, m_{D_s^*}} \, q^2 \nn \\
&  \times & \bigg\{ g_1^S \, [ m_{D_{s0}^*}^2 - m_{D_s^*}^2 - q^2 - 4 \, m_\ell^2 ] - \frac{4 \, g_2^S}{\Lambda} \, [ 2 \, m_{D_{s0}^*} \, m_\ell^2 - m_{D_s^*} \, ( q^2 + 2 \, m_\ell^2 ) ] \bigg\} \\
B_\text{int} (q^2)& = & - g(q^2) \, f(q^2) \, \tau_{1/2}(q^2) \, \frac{\lambda(m_{D_{s0}^*}^2, m_{D_s^*}^2, q^2)}{6 \, \sqrt{2} \, m_{D_{s0}^*}^3 \, m_{D_s^*}} \, ( q^2 - 4 \, m_\ell^2 ) \nn \\
&  \times & \bigg\{ g_1^S \, [ m_{D_{s0}^*}^2 - m_{D_s^*}^2 + q^2 ] + \frac{4 \, g_2^S}{\Lambda} \, m_{D_{s0}^*}\,q^2 \bigg\} \,\,\, .
\eea

\vspace{1cm}
\noindent {\fbox{$\bf D_{s1}^\prime\to D_s^* \ell^+  \ell^-$}}
\bea
A_c (q^2)& = & g(q^2) \, \big[ \tau_{1/2}(q^2) \big]^2 \, \frac{[ ( m_{D_{s1}^\prime} - m_{D_s^*} )^2 - q^2 ]^2}{9 \, m_{D_{s1}^\prime}^3 \, m_{D_s^*}} \, q^2 \, [ ( m_{D_{s1}^\prime} + m_{D_s^*} )^2 + q^2 + 4 \, m_\ell^2 ] \\
B_c (q^2)& = & - g(q^2) \, \big[ \tau_{1/2}(q^2) \big]^2 \, \frac{[ ( m_{D_{s1}^\prime} - m_{D_s^*} )^2 - q^2 ]^2}{9 \, m_{D_{s1}^\prime}^3 \, m_{D_s^*}} \, ( q^2 - 4 \, m_\ell^2 ) \, [ ( m_{D_{s1}^\prime} + m_{D_s^*} )^2 - q^2 ] \qquad \\
A_s (q^2)& = & g(q^2) \,  f^2(q^2)  \, \frac{[ ( m_{D_{s1}^\prime} + m_{D_s^*} )^2 - q^2 ]^2}{72 \, m_{D_{s1}^\prime}^3 \, m_{D_s^*}} \, q^2 \nn \\
& \times & \bigg\{ \big( g_1^S \big)^2 \, [ ( m_{D_{s1}^\prime} - m_{D_s^*} )^2 + q^2 + 4 \, m_\ell^2 ] + \frac{8 \, g_1^S \, g_2^S}{\Lambda} \, \big( m_{D_{s1}^\prime} - m_{D_s^*} \big) \, ( q^2 + 2 \, m_\ell^2 )  \nn \\
& + & \frac{4 \, (g_2^S)^2}{\Lambda^2} \, \big[ q^4 + q^2 \, ( m_{D_{s1}^\prime} - m_{D_s^*} )^2 + 4 \, m_\ell^2 \, ( m_{D_{s1}^\prime} - m_{D_s^*} )^2 \big] \bigg\} \\
B_s (q^2)& = & - g(q^2) \,  f^2(q^2)  \, \frac{[ ( m_{D_{s1}^\prime} + m_{D_s^*} )^2 - q^2 ]^2}{72 \, m_{D_{s1}^\prime}^3 \, m_{D_s^*}} \, [ ( m_{D_{s1}^\prime} - m_{D_s^*} )^2 - q^2 ] \, ( q^2 - 4 \, m_\ell^2 ) \nn \\
& \times& \bigg\{ \big( g_1^S \big)^2 - \frac{4 \, (g_2^S)^2}{\Lambda^2} \, q^2 \bigg\} \\
A_\text{int} (q^2)& = & g(q^2) \, f(q^2) \, \tau_{1/2}(q^2) \, \frac{\lambda(m_{D_{s1}^\prime}^2, m_{D_s^*}^2, q^2)}{9 \, \sqrt{2} \, m_{D_{s1}^\prime}^3 \, m_{D_s^*}} \, q^2 \, ( m_{D_{s1}^\prime} + m_{D_s^*} ) \nn \\
& \times & \bigg\{ g_1^S \, ( m_{D_{s1}^\prime} - m_{D_s^*} ) + \frac{2 \, g_2^S}{\Lambda} \, q^2 \bigg\} \\
B_\text{int} (q^2)& = & - g(q^2) \, f(q^2) \, \tau_{1/2}(q^2) \, \frac{\lambda(m_{D_{s1}^\prime}^2, m_{D_s^*}^2, q^2)}{9 \, \sqrt{2} \, m_{D_{s1}^\prime}^3 \, m_{D_s^*}} \, ( q^2 - 4 \, m_\ell^2 ) \, ( m_{D_{s1}^\prime} + m_{D_s^*} ) \nn \\
&  \times & \bigg\{ g_1^S \, ( m_{D_{s1}^\prime} - m_{D_s^*} ) + \frac{2 \, g_2^S}{\Lambda} \, q^2 \bigg\} \,\, . 
\eea

\vspace{1cm}
\noindent  {\fbox{$\bf D_{s1}\to D_s^*  \ell^+  \ell^-$}}
\begin{align}
A_c (q^2)& = g(q^2) \, \big[ \tau_{3/2}(q^2) \big]^2 \, \frac{ \lambda^2(m_{D_{s1}}^2, m_{D_s^*}^2, q^2) }{432 \, m_{D_{s1}}^5 \, m_{D_s^*}^3} \, q^2 \, [ 2 \, ( m_{D_{s1}} + m_{D_s^*} )^2 + 5 \,  ( q^2 + 4 \, m_\ell^2 ) ] \\
B_c (q^2)& = - g(q^2) \, \big[ \tau_{3/2}(q^2) \big]^2 \, \frac{ \lambda^2(m_{D_{s1}}^2, m_{D_s^*}^2, q^2) }{432 \, m_{D_{s1}}^5 \, m_{D_s^*}^3} \, ( q^2 - 4 \, m_\ell^2 ) \, [ 2 \, ( m_{D_{s1}} + m_{D_s^*} )^2 - 5 \, q^2 ] \\
A_s(q^2) & = g(q^2) \,  f^2(q^2)  \, \frac{h_T^2}{\Lambda^4} \, \frac{ \lambda^2(m_{D_{s1}}^2, m_{D_s^*}^2, q^2) }{864 \, m_{D_{s1}}^5 \, m_{D_s^*}} \, q^2 \, [ 2 \, q^4 + 5 \, ( q^2 + 4 \, m_\ell^2 ) \, ( m_{D_{s1}} + m_{D_s^*} )^2 ] \\
B_s (q^2)& = g(q^2) \,  f^2(q^2)  \, \frac{h_T^2}{\Lambda^4} \, \frac{\lambda^2(m_{D_{s1}}^2, m_{D_s^*}^2, q^2) }{864 \, m_{D_{s1}}^5 \, m_{D_s^*}} \, q^2 \, ( q^2 - 4 \, m_\ell^2 ) \, [ 5 \, ( m_{D_{s1}} + m_{D_s^*} )^2 - 2 \, q^2 ] \\
A_\text{int} (q^2)& = g(q^2) \, f(q^2) \, \tau_{3/2}(q^2) \, \frac{h_T}{\Lambda^2} \, \frac{ \lambda^2(m_{D_{s1}}^2, m_{D_s^*}^2, q^2) }{216 \, \sqrt{2} \, m_{D_{s1}}^5 \, m_{D_s^*}^2} \, q^2 \, ( q^2 + 12 \, m_\ell^2 ) \, ( m_{D_{s1}} + m_{D_s^*} ) \\
B_\text{int} (q^2)& = g(q^2) \, f(q^2) \, \tau_{3/2}(q^2) \, \frac{h_T}{\Lambda^2} \, \frac{5 \,  \lambda^2(m_{D_{s1}}^2, m_{D_s^*}^2, q^2) }{216 \, \sqrt{2} \, m_{D_{s1}}^5 \, m_{D_s^*}^2} \, q^2 \, ( q^2 - 4 \, m_\ell^2 ) \, ( m_{D_{s1}} + m_{D_s^*} ) \,\,\, .
\end{align}

\noindent {\fbox{$\bf D_{s2}^*\to D_s^*  \ell^+  \ell^-$}}
\begin{align}
A_c(q^2) & = g(q^2) \, \big[ \tau_{3/2}(q^2) \big]^2 \, \frac{ \lambda^2(m_{D_{s2}^*}^2, m_{D_s^*}^2, q^2) }{720 \, m_{D_{s2}^*}^5 \, m_{D_s^*}^3} \, q^2 \, [ 10 \, ( m_{D_{s2}^*} + m_{D_s} )^2 + 7 \, ( q^2 + 4 \, m_\ell^2 ) ] \\
B_c (q^2)& = - g(q^2) \, \big[ \tau_{3/2}(q^2) \big]^2 \, \frac{ \lambda^2(m_{D_{s2}^*}^2, m_{D_s^*}^2, q^2) }{720 \, m_{D_{s2}^*}^5 \, m_{D_s^*}^3} \, ( q^2 - 4 \, m_\ell^2 ) \, [ 10 \, ( m_{D_{s2}^*} + m_{D_s^*} )^2 - 7 \, q^2 ] \\
A_s (q^2)& = g(q^2) \, \big[ f(q^2) \big]^2 \, \frac{h_T^2}{\Lambda^4} \, \frac{ \lambda^2(m_{D_{s2}^*}^2, m_{D_s^*}^2, q^2) }{1440 \, m_{D_{s2}^*}^5 \, m_{D_s^*}} \, q^2 \, [ 10 \, q^4 + 7 \, ( q^2 + 4 \, m_\ell^2 ) \, ( m_{D_{s2}^*} + m_{D_s^*} )^2 ] \\
B_s (q^2)& = g(q^2) \, \big[ f(q^2) \big]^2 \, \frac{h_T^2}{\Lambda^4} \, \frac{ \lambda^2(m_{D_{s2}^*}^2, m_{D_s^*}^2, q^2) }{1440 \, m_{D_{s2}^*}^5 \, m_{D_s^*}} \, q^2 \, ( q^2 - 4 \, m_\ell^2 ) \, [ 7 \, ( m_{D_{s2}^*} + m_{D_s} )^2 - 10 \, q^2 ] \\
A_\text{int}(q^2) & = - g(q^2) \, f(q^2) \, \tau_{3/2}(q^2) \, \frac{h_T}{\Lambda^2} \, \frac{ \lambda^2(m_{D_{s2}^*}^2, m_{D_s^*}^2, q^2) }{360 \, \sqrt{2} \, m_{D_{s2}^*}^5 \, m_{D_s^*}^2} \, q^2 \, ( 17 \, q^2 + 28 \, m_\ell^2 ) \, ( m_{D_{s2}^*} + m_{D_s^*} ) \\
B_\text{int}(q^2)& = g(q^2) \, f(q^2) \, \tau_{3/2}(q^2) \, \frac{h_T}{\Lambda^2} \, \frac{ \lambda^2(m_{D_{s2}^*}^2, m_{D_s^*}^2, q^2) }{120 \, \sqrt{2} \, m_{D_{s2}^*}^5 \, m_{D_s^*}^2} \, q^2 \, ( q^2 - 4 \, m_\ell^2 ) \, ( m_{D_{s2}^*} + m_{D_s^*} ) \,\,\, .
\end{align}

\newpage
\bibliographystyle{JHEP}
\bibliography{refsDs}
\end{document}